\documentclass[letterpaper,twocolumn,10pt,anonymous]{article}
\usepackage{usenix2019_v3}

\usepackage{tikz}
\usepackage{amsmath}

\usepackage{subcaption}
\usepackage{graphicx}
\usepackage{multirow}

\usepackage{algorithm}
\usepackage{algpseudocode}
\usepackage{algorithmicx}
\algrenewcommand\algorithmicrequire{\textbf{Input:}}
\algrenewcommand\algorithmicensure{\textbf{Output:}}

\usepackage{listings}

\usepackage[compact]{titlesec}
\titlespacing*{\section}{0pt}{*0.8}{*0.8}
\usepackage{filecontents}

\begin{document}

\date{}

\title{\Large \bf MultiPaxos Made Complete}

\author{
{\rm Zhiying Liang}\\
The Pennsylvania State University\\
zvl5490@psu.edu
\and
{\rm Vahab Jabrayilov\thanks{Work was done as an intern at The Pennsylvania State University}} \\
Columbia University\\
vj2267@columbia.edu
\and
{\rm Aleksey Charapko}\\
University of New Hampshire\\
Aleksey.Charapko@unh.edu
\and
{\rm Abutalib Aghayev}\\
The Pennsylvania State University\\
agayev@psu.edu
} 

\maketitle

\begin{abstract}
MultiPaxos, while a fundamental Replicated State Machine algorithm, suffers from a dearth of comprehensive guidelines for achieving a complete and correct implementation.
This deficiency has hindered MultiPaxos' practical utility and adoption and has resulted in flawed claims about its capabilities.
Our paper aims to bridge the gap between MultiPaxos' complexity and practical implementation through a meticulous and detailed design process spanning more than a year. 
It carefully dissects each phase of MultiPaxos and offers detailed step-by-step pseudocode---in addition to a complete open-source implementation---for all components, including the leader election, the failure detector, and the commit phase. 

The implementation of our complete design also provides better performance stability, resource usage, and network partition tolerance than naive MultiPaxos versions. 
Our specification includes a lightweight log compaction approach that avoids taking repeated snapshots, significantly improving resource usage and performance stability. 
Our failure detector, integrated into the commit phase of the algorithm, uses variable and adaptive heartbeat intervals to settle on a better leader under partial connectivity and network partitions, improving liveness under such conditions. 

\end{abstract}

\section{Introduction}
State machine replication (SMR) is a cornerstone of modern fault-tolerant distributed computing services, such as cluster management and configuration, databases, and other data-intensive systems.
Utilizing replication strategies, SMR guarantees consistent command execution across peers, even during failures, by ensuring that every peer applies the same commands in an identical sequence to its state machine. 
While many SMR protocols exist, MultiPaxos consensus protocol remains the popular choice among major large-scale production systems such as Chubby~\cite{chubby}, Google Spanner~\cite{spanner}, Azure Storage~\cite{azurestorage}, and Amazon DynamoDB~\cite{dynamodb}. 

Despite over two decades of extensive discussion and optimization, translating its complex algorithm into practical, efficient, and scalable implementations remains a challenging task.
Previous studies have provided insights into their implementation experiences, yet these endeavors often fall short of providing a comprehensive and replicable blueprint for a correct and complete MultiPaxos-based application~\cite{paxos-made-live, paxos-made-practical, paxos-made-complex}. 
The advent of Raft in 2014, known for its well-defined specification, marked a significant shift in industry preference away from MultiPaxos~\cite{raft}.
This transition highlights an essential demand for clarity and simplicity in consensus protocols, especially as increasing distributed systems aim to achieve linearizable consistency~\cite{redis-raft, scylladb, facebook-mysql}. However, Raft may experience higher variance in the election time~\cite{paxos-vs-raft}, and its requirement of consecutive logs results in deadlock under some partial network partitions~\cite{omnipaxos}.

Ongoing research efforts in optimizing MultiPaxos primarily focus on enhancing performance and availability~\cite{epaxos, mencius, nopaxos, chain-paxos, linearizable-read, epaxos-revisited, vr, rabia}.
However, the absence of a standardized, correct, and detailed implementation poses the risk of misinterpreting how MultiPaxos works. 
This gap may result in missing key features like log trimming or lead to over-optimization, adding unnecessary complexity to issues that original MultiPaxos is capable of handling efficiently.

In this paper, we present a thorough design of MultiPaxos, accompanied by detailed pseudocodes provided in \autoref{multipaxos-psudocode} and \autoref{log-psudocode}, to bridge the existing gap between its theoretical algorithm and practical implementations.
Our work addresses and improves critical areas that lack clear implementation specifications in prior studies~\cite{paxos-made-live, paxos-made-practical, paxos-made-complex}, such as the prepare and commit phases and the failure detector.
We also tackle the challenge of effectively coordinating these complex components.
Beyond simply incorporating critical system features like log trimming and recovery, our design proposes a more streamlined approach that minimizes message complexity while ensuring readability and clarity.

In addition to the core aspects of MultiPaxos, our paper delves into further optimizations in log compaction. 
Traditional snapshotting methods~\cite{etcd, tikv, etcd-raft-example, hashicorp, redis-raft, braft}, though common for log compaction, often lead to notable performance drawbacks and are challenging to implement. 
Our exploration suggests that an alternative pathway where log compaction is achievable without snapshots. 
Specifically, once a command is applied to the state machine by a peer, it can be safely removed on that peer, irrespective of its status on other peers. The command has already been stored in the majority of peers' state machines or logs. 
However, it comes at a huge potential cost. It is possible that some peers miss one or more commands during the regular replication. When commands are trimmed too quickly, we need to copy the whole state machine for a single command recovery. 
Therefore, we propose a new adaptive and lightweight log compaction approach that avoids snapshotting and reduces the need for replicating the entire state machine.
This approach involves the leader broadcasting and collects the lowest index of the last executed instance in the log among all peers via the commit phase.
By iteratively computing a new global lowest index, peers autonomously and efficiently manage the removal of executed log entries.

Furthermore, we address the challenges posed by partial network partitions. We specifically focus on two types of network partitions. 
One scenario involves all servers being disconnected from each other but maintaining only a single connection to the same server, while the other involves two servers that are unreachable from each other but connected by a third server, a situation notably observed in the Cloudflare incident in 2020~\cite{cloudflare}. 
This partition leads to continuous leader elections, impeding progress due to the absence of stable leadership. 
To address it, we introduce a modification to the failure detector---an adaptive timeout mechanism.
This mechanism forces peers who excessively participate in leader elections within a specific timeframe to increment their timeout, delaying subsequent elections.
Eventually, a peer with a stable connection becomes the leader due to its shorter timeout.
Importantly, our mechanism enhances performance without altering the core design principles of MultiPaxos.

We highlight our key contributions as follows:
\begin{itemize}
\item We present comprehensive, complete, and understandable MultiPaxos design and implementation with detailed pseudocode.
\item We specify the coordination of the Prepare and Commit phases of MultiPaxos, outlining their expanded role in leader election, commit messages, and the failure detector. We also provide a detailed approach to log recovery for multiple instances in the Prepare phase.
\item We introduce a lightweight yet effective log compaction mechanism for MultiPaxos, designed to eliminate the need for continuous snapshotting.
\item We propose an adaptive timeout mechanism that enhances the resilience of MultiPaxos, particularly in scenarios involving partial network partition.
\end{itemize}
\section{Background}
\subsection{Single-Decree Paxos and MultiPaxos}
Single-Decree Paxos~\cite{parliament, paxos-made-simple}, a fundamental algorithm for solving distributed consensus, relies on a group of peers collectively agreeing on some non-trivial single value. 
The algorithm operates in two distinct phases: \textit{Prepare} and \textit{Accept}. 
In the \textit{Prepare Phase}, a peer prepares to become a leader with a unique ballot number higher than any it has encountered. If, upon receiving promises from a majority of peers, the prospective leader concludes that the majority has not seen a higher ballot than its own one, then it becomes a leader. 
Subsequently, the leader selects a value with the highest ballot among the received promises or a new value if no values are found in the promises. 
Moving forward, the leader advances to the \textit{Accept Phase} where peers accept the proposed value. 
Once the leader receives a majority of acknowledgments, the consensus is reached. At this point, it can dispatch commit messages to let all peers learn about the agreed value.

\begin{figure}
    \centering
    \includegraphics[width=\linewidth]{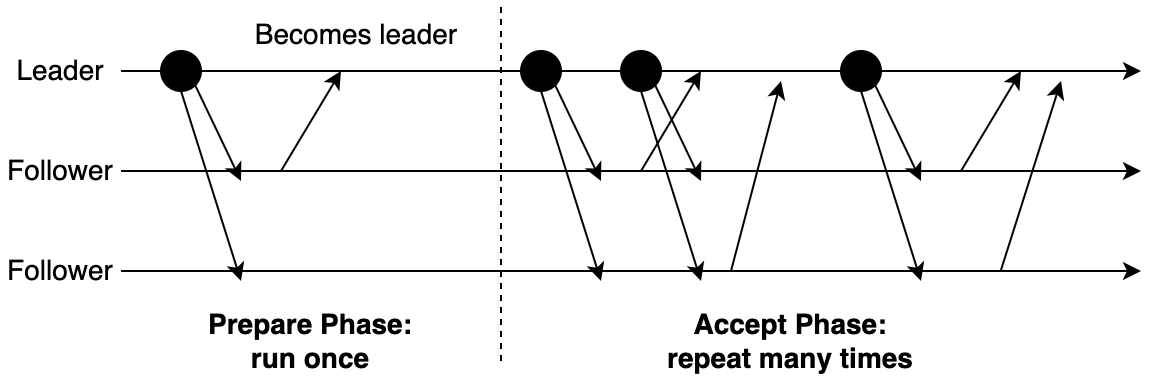}
    \caption{Overview of the MultiPaxos Algorithm}
    \label{fig:multipaxos-overview}
\end{figure}

While Single-Decree Paxos is well-specified and easy to implement, it is not widely used in practical systems, as it requires multiple message round-trips for each value. 
MultiPaxos, a well-known optimization, reduces unnecessary messages. 
\autoref{fig:multipaxos-overview} illustrates that, in MultiPaxos, once the leader is elected, it gains the ability to directly propose values for an unbounded number of instances during the \textit{Accept Phase}, thus eliminating the need for the repeated \textit{Prepare} messages. 
The \textit{Prepare Phase} only comes into play in the event of a leader crash. 

\subsection{Identifying the Specification Gaps}
MultiPaxos has been extensively discussed in Lamport’s Paxos papers and its subsequent optimizations~\cite{mencius, flexible-paxos, epaxos, nopaxos, paxos-made-live, paxos-made-practical, epaxos-revisited}. 
However, in contrast to Single-Decree Paxos, significant gaps still exist between the algorithmic description of MultiPaxos and its practical implementation.
These gaps, listed below, span the entire MultiPaxos algorithm.

\textbf{Leader Election}. The \textit{Prepare Phase} is a crucial phase of MultiPaxos, but the algorithm leaves many details unclear, such as how to record and track leadership, when to update the ballot number, and how to handle log recovery.
While using a separate field to track leadership and having an independent log recovery phase is a common approach, it notably adds complexity to the implementation~\cite{nopaxos, omnipaxos, skyros}.

\textbf{Commit Phase}. This phase involves broadcasting committed messages to enable followers to synchronize with the leader's progress. 
While conceptually straightforward, its overhead in practical applications is non-trivial. 
Having individual commit messages for each log instance could potentially double the message traffic within the cluster~\cite{paxi, franken-paxos, epaxos, epaxos-revisited}, while batch committing requires a well-designed method to manage the commit rate and the size of messages.

\textbf{Failure Detector}. Once a leader is elected, it needs to broadcast its leadership, and followers rely on a failure detector to monitor the leader’s status. 
However, there is a lack of detailed guidelines for integrating MultiPaxos seamlessly with the failure detector. 
Most implementations employ the timeout mechanism. Unlike in raft, if heartbeat messages in MultiPaxos do not carry additional information, it simply causes an increase in network traffic without benefits.

\textbf{Log Compaction}. Log compaction is essential to prevent unbounded growth of the log. While periodic snapshotting is widely used, it needs to copy the state machine exclusively, thus resulting in performance degradation periodically. Furthermore, it requires a complex design, involving disk I/O and the management of snapshot files.

\textbf{Partial Network Partition}. While partial partitions do not violate any safety in the MultiPaxos Algorithm, they can significantly reduce MultiPaxos's availability in practice. It is important to figure out what partition scenarios MultiPaxos can handle and what situations would lead to livelock. Enhancing resilience under such conditions is crucial for practical deployment.

\section{Bridging the Specification Gaps}

\begin{algorithm}[t]
\caption{Prepare Phase and its handler}
\label{alg:prepare}
\centering
\textbf{Prepare Thread}
\begin{algorithmic}[1]
\State Sleep until the peer becomes a follower.
\While{$commit\_received == true$}
    \State $commit\_received \leftarrow false$; sleep between 2.5-3x $commit\_interval$ (\S\ref{failure-detector}).
\EndWhile
\State $PBN \leftarrow ABN +$ constant\_round\_increment.
\State Broadcast Prepare requests with $PBN$
\While{Receive a reply}
    \If{$reply.type == OK$}
        \State Increment quorum count; merge $reply.instances$.
        \If{QUORUM} 
            \State $ABN \leftarrow PBN$, re-run \textit{Accept} for the merged log; GOTO step 1.
        \EndIf
    \Else
        \State $ABN \leftarrow reply.ballot$; GOTO step 2.
    \EndIf
\EndWhile
\State GOTO step 2 due to insufficient replies.
\end{algorithmic}

\bigskip

\textbf{Prepare Handler}
\begin{algorithmic}[1]
\Require \underline{request\_ballot}: sender's ballot;
\Ensure \underline{type}: OK or REJECT; 
\underline{ballot}: $ABN$ of the peer;
\underline{instances}: all existing log instances
\State If $request\_ballot > ABN$, $ABN \leftarrow request\_ballot$, awake \textit{Prepare Thread}, and reply OK with all existing instances in its log.
\State Else, reply REJECT with its current $ABN$.
\end{algorithmic}
\end{algorithm}

We now present the design of our MultiPaxos module\footnote{\url{https://github.com/psu-csl/replicated-store}} and highlight our decisions made to bridge the specification gaps of MultiPaxos. 

To enhance both the practicality and simplicity of our MultiPaxos implementation, we introduce supplementary components beyond the scope of MultiPaxos, extending it to a MultiPaxos-based linearizable distributed key-value store. 
Within this application, the MultiPaxos module plays a central role---responsible for MultiPaxos protocol and inter-peer communication. 
It runs along with the log module, which provides a thread-safe unbounded producer-consumer queue. The log module maintains the order of commands that will be applied to the state machine. 
All modules are highly decoupled, with the MultiPaxos module also providing interfaces for the upper application layer.

In this section, we focus on how our MultiPaxos design addresses specific gaps. 
We begin by clarifying the unclear specifications for leader election, as well as the accept and commit phase, which form the core sequence of the MultiPaxos algorithm. 
Following this, we then address how to integrate the failure detector, perform log compaction, and enhance resilience against partial network partitions.

\subsection{Leader Election}
\label{leader-election}

A pivotal distinction between Single-Decree Paxos and our MultiPaxos implementation is the leadership model. 
In MultiPaxos, a specific peer is designated as the long-term leader, thereby skipping numerous redundant \textit{Prepare} requests. 
Once elected, the leader can issue Accept requests directly, bypassing the \textit{Prepare Phase}. 
In this section, we aim to bridge the gap in the leader election process by detailing its implementation specifications.

To implement the leader election module, we assign each peer one of two roles: \textit{Leader} or \textit{Follower}. Peers use the Active Ballot Number ($ABN$) to track leadership.
Additionally, we employ a long-running thread---\textit{Prepare Thread}. 
Algorithm \autoref{alg:prepare} shows that the \textit{Prepare Thread} primarily determines when to initialize a new leader election (lines 2-4) and performs the leader election (lines 5-17). 
Initially, each peer runs as a \textit{Follower}, with the \textit{Prepare Thread} activated immediately at the beginning. 
In the \textit{Follower} role, the \textit{Prepare Thread} intermittently sleeps for a random period, then assesses the status of its failure detector to determine whether to initiate a leader election. 

\begin{figure}[t]
    \centering
    \includegraphics[width=\linewidth]{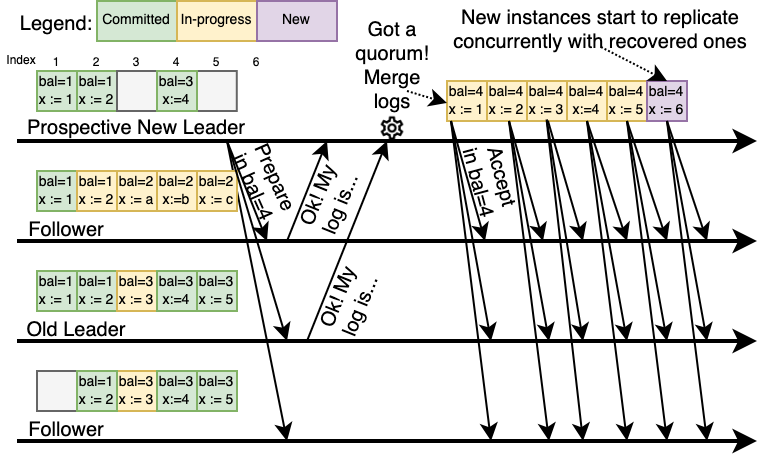}
    \caption{An example of the \textit{Prepare} (left) and \textit{Accept} (right) Phase. Each peer maintains a log consisting of instances. An instance contains a ballot (denoted as 'bal') and a value (e.g., x:=1). Initially, instances are marked as \textit{In-progress} when the leader adds them to the log. They transition to \textit{Committed} when safe for execution. The 'New' state, while not an actual state, is used to distinguish between new instances and existing ones. Note that, in the \textit{Accept Phase}, the figure only depicts the merged log instead of the actual log to save space.}
    \label{fig:prepare-phase}
\end{figure}

Upon the initiation of a leader election, the \textit{Prepare Thread} creates a ballot, namely the Prepare Ballot Number ($PBN$), higher than $ABN$. 
Each peer possesses a distinct range of available ballot options, denoted as $peer\_id + round * max\_num\_peers$, ensuring the uniqueness of each ballot. 
Similar to Single-Decree Paxos, the \textit{Prepare Thread} dispatches \textit{Prepare} requests with the ballot number and awaits replies until it reaches the quorum. 
A reply has two types: \textit{OK} and \textit{REJECT}. Only \textit{OK} replies contribute to the quorum. 
While receiving a \textit{REJECT} reply, the peer learns a new leader by updating its $ABN$, shown in Algorithm \autoref{alg:prepare} (lines 8-14). We adopt this approach for all types of replies in our MultiPaxos design.
Subsequently, \textit{Prepare Thread}, once the peer attains leadership, enters a sleep state on a condition variable, awakening only when the peer relinquishes leadership.

Different from previous studies~\cite{paxos-made-complex, paxi}, a notable decision is that we cache the ballot for elections in $PBN$ and defer the update to $ABN$ until the successful completion of the leader election.
We ensure that $ABN$ is updated only when the actual leadership is changed, either becoming a leader or encountering a new leader.
If we update the $ABN$ at the onset of the leader election, we require additional leadership flags to help the peer distinguish between its current ballot number and actual leadership. 
Otherwise, it may proceed to the \textit{Accept Phase} wrongly, as $ABN$ is updated before the peer becomes a leader.
Such two-leadership-flag approach can be found in previous studies~\cite{paxos-made-complex}. In actual implementations, it usually requires an explicit lock to update both fields at the same time.
Our approach, however, eliminates the need for locking and redundant comparisons of ballot numbers, such as double-checking leadership status at the end of the leader election, thus mitigating the associated overhead of atomic operations.   

In addition to fulfilling the leader election, log recovery is also essential in MultiPaxos.
One of the major differences between MultiPaxos and Raft is that a sequential and up-to-date log is not a prerequisite for a peer to become a leader in MultiPaxos. 
Thus, a new leader may not have a complete log, and any gaps in the leader's log may cause progression stalls.
We address this by efficiently orchestrating log recovery during the \textit{Prepare Phase}. 
As shown in Algorithm \autoref{alg:prepare}, in the \textit{Prepare} request handler, when a peer promises a \textit{Prepare} request, it responds with not only an OK reply but also all existing instances in its log. 
Line 9 in Algorithm \autoref{alg:prepare} shows that, the \textit{Prepare Thread} merges all instances from the response and caches them in a temporary log when it receives a vote. 
Each instance in this temporary log can be in one of the three states: \textit{Executed} (applied to the state machine), \textit{Committed} (quorum reached), and \textit{In-progress} (initial unsafe state).

For example, as illustrated in \autoref{fig:prepare-phase}, a new leader is unaware of certain log instances: index 3 and 5. 
By collecting the majority of \textit{Promise} responses, the new leader learns all existing instances from other peers, as all \textit{Committed} instances are stored on the majority of peers.
In instances where conflicts arise, such as index 3, the leader selects the instance with the higher ballot number. 
After merging all existing instances from these responses, the peer officially assumes leadership and re-runs replication (\textit{Accept Phase}) for the merged log (namely replay), enabling followers to recover their missing instances individually as well.

\begin{algorithm}[t]
\caption{Accept Phase and its handler}
\label{alg:accept}
\centering
\textbf{Leader Accept Phase}
\begin{algorithmic}[1] 
\State Extract $leader\_id$ from $ABN$.
\If{$peer.id == leader\_id$}
    \State Create an instance and append it to the log.
\ElsIf{$leader\_id ==$ other peers'id}
    \State Return $leader\_id$.
\Else
    \State Return retry (the initial leader election is in progress).
\EndIf
\State Broadcast \textit{Accept} requests with instance and $ABN$.
\While{Receive a reply}
    \If{$reply.type == OK$}
        \State Increment quorum count.
        \If{QUORUM} 
            \State Commit the instance and return.
        \EndIf
    \Else
        \State $ABN\leftarrow reply.ballot$; extract, return $leader\_id$
    \EndIf
\EndWhile
\State Return retry.
\end{algorithmic}

\bigskip

\textbf{Follower's Accept Handler}
\begin{algorithmic}[1]
\Require \underline{request\_ballot}: ballot of the instance;
\underline{instance}: wrapper of the command;
\Ensure \underline{type}: OK or REJECT;
\underline{ballot}: ABN of the peer
\State If $request\_ballot \geq ABN$, append the instance to the log.
\State If $request\_ballot > ABN$, $ABN \leftarrow request\_ballot$ and awake \textit{Prepare Thread}.
\State If $request\_ballot < ABN$, reply REJECT with its current $ABN$ and return.
\State Reply OK.
\end{algorithmic}
\end{algorithm}

\subsection{Accept Phase}

Once the leadership is established, the leader is ready to receive client requests and initiate the replication phase, notably the \textit{Accept Phase}. 
Lines 2-9 of the leader part in Algorithm \autoref{alg:accept} shows how it works. The leader generates an instance for each client command, adds it to its log, and then broadcasts the instance to all followers, accompanying it with its current $ABN$.
When the follower receives an \textit{Accept} request, it simply compares the ballot number and decides whether to add the instance in its log, as shown in the follower's handler in Algorithm \autoref{alg:accept}.
While it is akin to the approach of Single-Decree Paxos, we have implemented several optimizations detailed below to enhance availability and parallelism.

A key feature of our design is that the leader begins the \textit{Accept Phase} immediately after setting the $ABN$ and runs it concurrently with the replay of the merged log from the Prepare Phase. 
Despite log recovery being an integral part of the \textit{Prepare Phase}, our exploration suggests that there is no conflict between it and the Accept Phase.
Our concurrent approach allows the leader to efficiently re-dispatch existing instances while simultaneously handling new client requests.  
\autoref{fig:prepare-phase} shows that the leader re-runs \textit{the Accept Phase} for Index 1-5 and replicates the new instance (denoted as 'new').
Such parallelism is crucial for maintaining cluster availability, ensuring no delays in the \textit{Accept Phase}, particularly when replaying a large number of instances.

Furthermore, another distinctive aspect of our design is that the leader can handle multiple replications concurrently.
For instance, with 64 outstanding close-loop clients, the leader utilizes 64 handlers to operate the replication process simultaneously.
In contrast, many other implementations employ a buffer to store client requests, and the leader replicates one request at a time~\cite{paxi,epaxos,skyros,libpaxos}. It often leads to a significant increase in latency. 
Our approach maintains low latency while effectively leveraging the parallelism within the system.

\subsection{Commit Phase with Heartbeats}
\label{commit-phase}

\begin{figure}
    \centering
    \includegraphics[width=\linewidth]{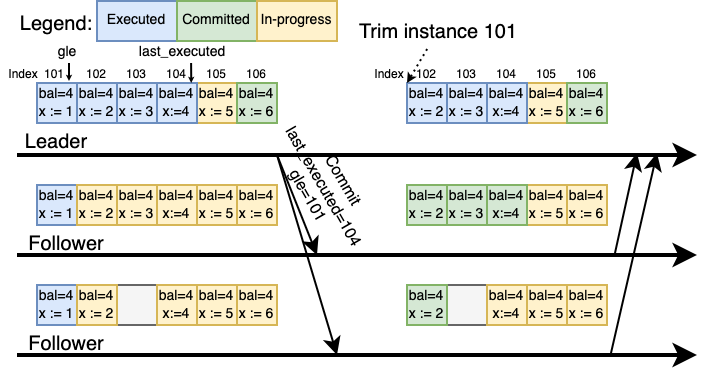}
    \caption{An example of the Commit Phase. The left part depicts the logs of all peers before the leader sends a new round of heartbeats. \textit{last\_executed} indicates the index of the last executed instance, and \textit{gle} refers to the global minimum of \textit{last\_executed} across peers. The right section shows log changes after the leader's heartbeats and before followers' responses. In addition to \textit{Committed} and \textit{In-progress}, \textit{Executed} represents instances already applied to the state machine.}
    \label{fig:commit-phase}
\end{figure}

Following the \textit{Accept Phase}, a commit message is necessary to enable followers to commit and apply decided instances to their state machines respectively. 
While having a distinct commit message for each \textit{Committed} instance seems a straightforward and simple option, it could result in an influx of requests and increased network traffic. One potential optimization is to piggyback committed messages onto the next \textit{Accept} requests~\cite{pigpaxos}. 
However, this method is contingent on the frequency of Accept requests, which is unstable and closely tied to the number of client requests. 
In scenarios where Accept requests are sparse, the leader might still need to send additional commit messages.

To address this issue, we propose to send a periodic request that contains information for batch committing, namely \textit{Commit} request, to obviate the need for multiple individual commit messages. 
Although the \textit{Commit} request conveys information for batch commits, it does not encompass every committed index on the leader’s side. 
Instead, it includes only one value---the index of the last executed instance (\textit{\textbf{last\_executed}}) of the leader, as shown in line 5 of the leader part in Algorithm \autoref{alg:commit}. 
This design minimizes the request size, keeping it consistently low and bounded. 
Line 1 of the follower part in Algorithm \autoref{alg:commit} shows that when the follower receives the \textit{Commit} request, it commits sequential instances until it reaches the index of \textit{\textbf{last\_executed}} provided by the leader or an absent instance.

\begin{algorithm}[t]
\caption{Commit Phase and its handler}
\label{alg:commit}
\centering
\textbf{Leader Commit Thread} \\
\begin{algorithmic}[1]
\State Sleep until the peer becomes a leader.
\State $gle \leftarrow$ global\_last\_executed read from the log (\S\ref{log-compaction}).
\While{Being Leader}
    \State Trim its log up to $gle$ (\S\ref{log-compaction}).
    \State Broadcast \textit{Commit} with $ABN$, $gle$, $last\_executed$.
    \While{Receive a reply}
        \If{$reply.type == OK$}
            \State Increment ok\_replies count.
            \If{Recv all follower's replies} 
                \State $gle \leftarrow$ min(replies' $last\_executed$) (\S\ref{log-compaction}).
            \EndIf
        \Else
            \State $ABN \leftarrow reply.ballot$; GOTO step 1.
        \EndIf
    \EndWhile
    \State Sleep for $commit\_interval$.
\EndWhile
\end{algorithmic}

\bigskip

\textbf{Follower's Commit Handler}
\begin{algorithmic}[1]
\Require \underline{request\_ballot}: ballot of the sender;
\underline{last\_executed}: last executed of the sender;
\underline{gle}: global\_last\_executed of the sender;
\Ensure \underline{type}: OK or REJECT;
\underline{ballot}: ballot of the peer;
\underline{own\_last\_executed}: last executed of the follower;
\State If $request\_ballot \geq ABN$, $commit\_received \leftarrow true$ (\S\ref{failure-detector}), commit sequential instances up to $last\_executed$ (\S\ref{commit-phase}), and trim the log up to $gle$(\S\ref{log-compaction});
\State If $request\_ballot > ABN$, $ABN \leftarrow request\_ballot$, and awake \textit{Prepare Thread}.
\State If $request\_ballot < ABN$, reply REJECT with its current $ABN$ and return;
\State Reply OK with its $last\_executed$.
\end{algorithmic}
\end{algorithm}

\autoref{fig:commit-phase} illustrates how the \textit{Commit} phase works in our design. 
When the leader sends a Commit request, it includes the index 104 as its \textit{\textbf{last\_executed}}.  
Once followers receive the request from the leader, they commit instances up to index 104. 
If there are missed instances before the \textit{\textbf{last\_executed}} sent by the leader, such as with Follower 2, the follower simply pauses the commit process. 
Although all existing instances between the gap and \textit{\textbf{last\_executed}} are deemed safe to commit, such as instance 104 in this case, they cannot be applied to the state machine, regardless of their states. Thus, we leave these instances uncommitted.

It is important to note that Paxos does not require sequential instances in the log. 
Sending the index of the last committed instance rather than \textit{\textbf{last\_executed}} could lead to ambiguity, as some instances between these two indices remain in the \textit{In-progress} state. 
For instance, if index 105 is still in the \textit{In-progress} state, followers would not have sufficient information to determine which instances are safe to commit. 
Consequently, instance 105 may be wrongly committed on the follower's side ahead of the leader. 
To avoid such scenarios, we opt for the \textit{\textbf{last\_executed}} index for our Commit requests. This method makes a trade-off between simplifying the commit process and the commit speed of followers by using \textit{\textbf{last\_executed}}.

The potential downside of this solution is the execution delay at the followers when the interval between \textit{Commit} requests, denoted as \textit{\textbf{commit\_interval}} (CI), is large. 
However, since the leader commits immediately upon reaching the quorum and promptly responds to clients without waiting for the execution of followers, a minor delay in the followers’ progress does not impact the client-perceived latency. 

\subsection{Failure Detector}
\label{failure-detector}

\begin{table*}[htb]
\centering
\begin{tabular}{|c||c|c|c|}
\hline
 & \textbf{Common Path} & \textbf{Adding New Peers} & \textbf{Peers Re-connection} \\ \hline \hline
\textbf{Our Adaptive} & \multirow{2}{*}{No performance degradation.} & \multirow{2}{*}{Need a snapshot on demand.} & Compact pauses during disconnection;\\ \textbf{Log Trim} & & & No snapshots needed \\ 
\hline

\textbf{Periodic} & Snapshots are not useful; & \multirow{2}{*}{Use the most recent snapshot;} & \multirow{2}{*}{Use the most recent snapshot;} \\ 
\multirow{2}{*}{\textbf{Snapshotting}} & but stalling progression; & & \\ 
& causing higher disk I/O. & No extra overhead. & No extra overhead. \\ \hline

\multirow{2}{*}{\textbf{Quorum Trim}} & \multirow{2}{*}{No performance degradation.} & Need at least one & Need at least one \\ 
& & snapshot on demand. & snapshot on demand. \\ 
\hline
\end{tabular}
\caption{Comparison of compaction approaches in different scenarios.}
\label{tab:log-compaction}
\end{table*}

We next clarify how to integrate a failure detector with MultiPaxos, addressing another gap that is briefly specified in the existing literature~\cite{paxos-made-simple, paxos-made-live}. 
The failure detector assumes two main responsibilities: preserving leadership stability and initiating leader elections when necessary. 
Firstly, maintaining leadership is essential to prevent the system's performance from being disrupted by frequent and unnecessary leader elections. 
Secondly, as mentioned in section \ref{leader-election}, the failure detector also plays a pivotal role in deciding when to initiate these leader elections. Once the leader becomes non-functional, the failure detector signals followers to initiate a leader election and replace the old leader, ensuring the system’s availability. 

As the leader already sends the \textit{Commit} request periodically, we utilize \textit{Commit} requests as heartbeat messages and employ a timeout mechanism for triggering elections. 
This approach reduces the number of messages the leader sends. Specifically, the leader dispatches heartbeats every \textit{\textbf{commit\_interval}} duration, while followers monitor their arrival at intervals and start leader elections if they do not receive these heartbeats in time.

Similar to the \textit{Prepare Phase}, we have a long-running thread, namely \textit{Commit Thread}, to assume the job of maintaining leadership.
In contrast to the \textit{Prepare Thread}, the \textit{Commit Thread}sleeps on a condition variable when the peer runs as a follower. 
It becomes active only when assuming the role of a leader, starting to dispatch \textit{Commit} requests periodically, as illustrated in the leader’s lines 1-5 in Algorithm \autoref{alg:commit}.

On the follower’s side, followers monitor the state of the last received heartbeats---an atomic boolean variable (\textit{commit\_received})---associated with a random timer in the \textit{Prepare Thread}. 
Line 1 of the follower part in Algorithm \autoref{alg:commit} shows that, whenever the follower receives a \textit{Commit} request from the leader, it sets \textit{commit\_received} to ‘true’. 
Upon triggering a timeout by the random timer, it first checks the state of \textit{commit\_received} and then resets it to 'false', as shown in lines 2-4 of the Prepare Thread in Algorithm \autoref{alg:prepare}. 
A 'true' state indicates that the follower has successfully received heartbeats from the leader, while a 'false' state indicates a failure in heartbeat reception, prompting the follower to initiate a leader election process. 
We calibrate the timer to a specific range, typically 2 to 2.5 times the \textit{\textbf{commit\_interval}}, to avoid the simultaneous initiation of multiple leader elections~\cite{raft}.




\subsection{Adaptive Log Compaction}
\label{log-compaction}

As the log accumulates an increasing number of \textit{Executed} instances, log compaction becomes essential to prevent unbounded growth. 
While academic studies often either omit log compaction or propose impractical approaches~\cite{paxi, skyros, nopaxos, epaxos}, many industrial implementations simply rely on snapshotting~\cite{etcd, tikv, redis, hashicorp}. 
However, periodic snapshotting demands exclusive access to the state machine, leading to low availability during the compaction, especially when the size of the state machine is large. 
Thus, we propose an adaptive and lightweight log compaction approach that avoids snapshotting in most cases. 
More importantly, our exploration reveals that snapshotting is not an absolute prerequisite for log compaction, a point often neglected in previous works.

In an ideal scenario, MultiPaxos can have each peer delete an instance from the log immediately after applying it to the state machine (namely quorum-trim), without violating safety, where the instance has been committed by the quorum~\cite{paxos-made-complex, paxi}. 
Once an instance is applied to by the peer, the instance becomes redundant for that peer. 
In cases where a lagging peer attempts to recover an instance that has been removed by others, it can still catch up on the progress by requesting a copy of the entire state machine from other peers. 
As long as the majority of peers remain functional, there is no risk of losing any committed states. 

However, this quorum-trim approach is not practical for real-world deployment, as it may generate multiple snapshots as long as a peer lags. 
For example, when a peer misses the replication of an instance from the leader due to the network fluctuation, it has no way to recover this missing instance from the log on other peers directly, because this instance is committed by the majority and safely removed. 
This situation eventually results in copying the entire state machine, even though the peer missed one instance only.

Thus, we make a balance between maximizing the efficiency of log compaction and minimizing the likelihood of snapshot creation.
Specifically, in our design, each follower responds to the leader’s \textit{Commit} message with their local \textit{\textbf{last\_executed}} value. 
Upon receiving responses from \textbf{all} peers, the leader computes \textit{\textbf{global\_last\_executed}} (\textit{\textbf{gle}})---the minimum of all \textit{\textbf{last\_executed}} values---and includes it in subsequent \textit{Commit} message, illustrated in Algorithm \autoref{alg:commit}.
This enables all peers to safely trim their logs up to this \textit{\textbf{gle}} value. 
For example, as shown in the \autoref{fig:commit-phase}, the leader calculates the \textit{\textbf{gle}} of 101 based on the \textit{\textbf{last\_executed}} values across peers. When a follower receives a \textit{Commit} message with \textit{\textbf{a gle}} of 101, it trims all prior instances up to index 101.
The \textit{\textbf{global\_last\_executed}} ensures that all instances trimmed from the log have been applied to the state machine across all peers. 
Even if a different peer becomes a leader, they do not need instances before \textit{\textbf{global\_last\_executed}}.
This maximizes the efficiency of instance trimming and eliminates the need of transferring snapshots.

Furthermore, our log compaction approach, which dynamically adjusts the compaction rate during runtime, contrasts with other methods that requires pre-configured frequencies for compaction.
In each round of the \textit{Commit} message, the leader calculates the compaction rate from the \textit{Commit} responses of all followers autonomously. 
In the scenario where all peers work at a similar rate, our approach maintains fewer entries in the log. 
Conversely, when certain peers lag, it provides the leader with information about which peer experiences delays. 
This dynamic mechanism obviates the need for manual fine-tuning of configurations.

\autoref{tab:log-compaction} compares how all three log-compaction approaches behave under various scenarios: when all peers are functional and responsive (common path), when a new peer joins the cluster, and when a peer reconnects after a disconnection. 
Our compaction approach minimizes the need to take snapshots in most cases, thus reducing its impact on performance during compaction. Conversely, periodic snapshotting generates snapshots among all scenarios, while quorum trim also demands snapshots in the event of adding new peers or peers reconnecting.

The only downside of our approach occurs when any of the peers, say P, temporarily disconnects and misses a command in its log, shown in \autoref{tab:log-compaction}.
P’s \textit{\textbf{last\_executed}} and \textit{\textbf{global\_last\_executed}} stop advancing because commands after the gap cannot be executed until the missing command is recovered and applied. 
This leads to log growth across all peers until the missing commands are recovered.
A practical solution to this---and some other problems caused by poor network connectivity---is to have the existing leader periodically run a complete \textit{Prepare Phase}, as is done in Chubby\cite{chubby}, to fill the gaps in the logs of temporarily disconnected peers and resume the compaction.

\subsection{Partial Network Partition}

\begin{figure}[tb]
    \centering
    \includegraphics[width=\linewidth]{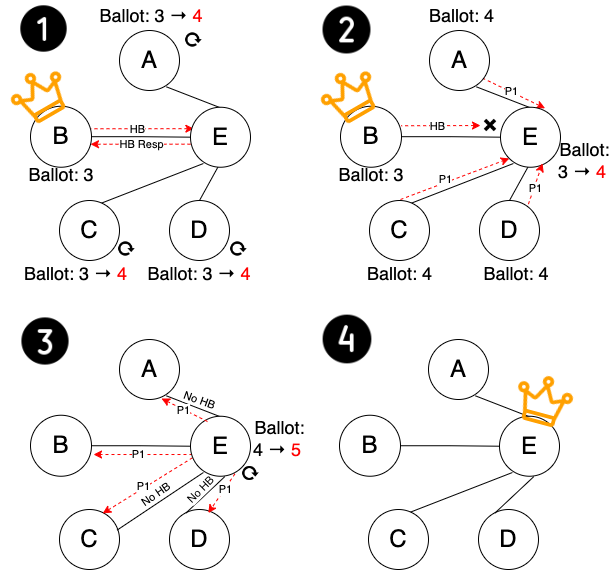}
    \caption{The leader-losing-quorum scenario. 1. Peer A, B, C, and D disconnect from each other but Peer E. E still answers the heartbeat from Leader B. 2. Peer A, C, and D trigger leader election but lack of enough votes to become a leader. But a vote with a higher term number trigger Peer E to ignore heartbeats from old leader B. 3. A will start election as it does not receives any heartbeats. 4. A becomes the leader, and the availability resumes.}
    \label{fig:leader-loss-quorum-example}
\end{figure}

Moving from normal to unusual network conditions, we next demonstrate the resilience of MultiPaxos and our enhancements, in scenarios of partial network partitions. 
A partial network partition is defined as a scenario where at least one peer loses connections with some peers, yet both of them remain connected by a third peer~\cite{nifty}. 
This situation poses a risk of livelocks in SMR systems, including MultiPaxos and Raft. 
A notable example is the outage failure within Cloudflare~\cite{cloudflare}. Prior studies summarize and analyze several types of partial partition scenarios~\cite{omnipaxos}. 
In contrast to the previous claim that MultiPaxos exhibits limited resilience in such situations~\cite{omnipaxos}, our key observation suggests that a proper failure detector design can adequately guarantee resilience without additional overhead. 
To validate this claim, we examine two representative types of partial partitions.

\subsubsection{Leader-Losing-Quorum Partition}

\begin{figure}[t]
    \centering
    \includegraphics[width=\linewidth]{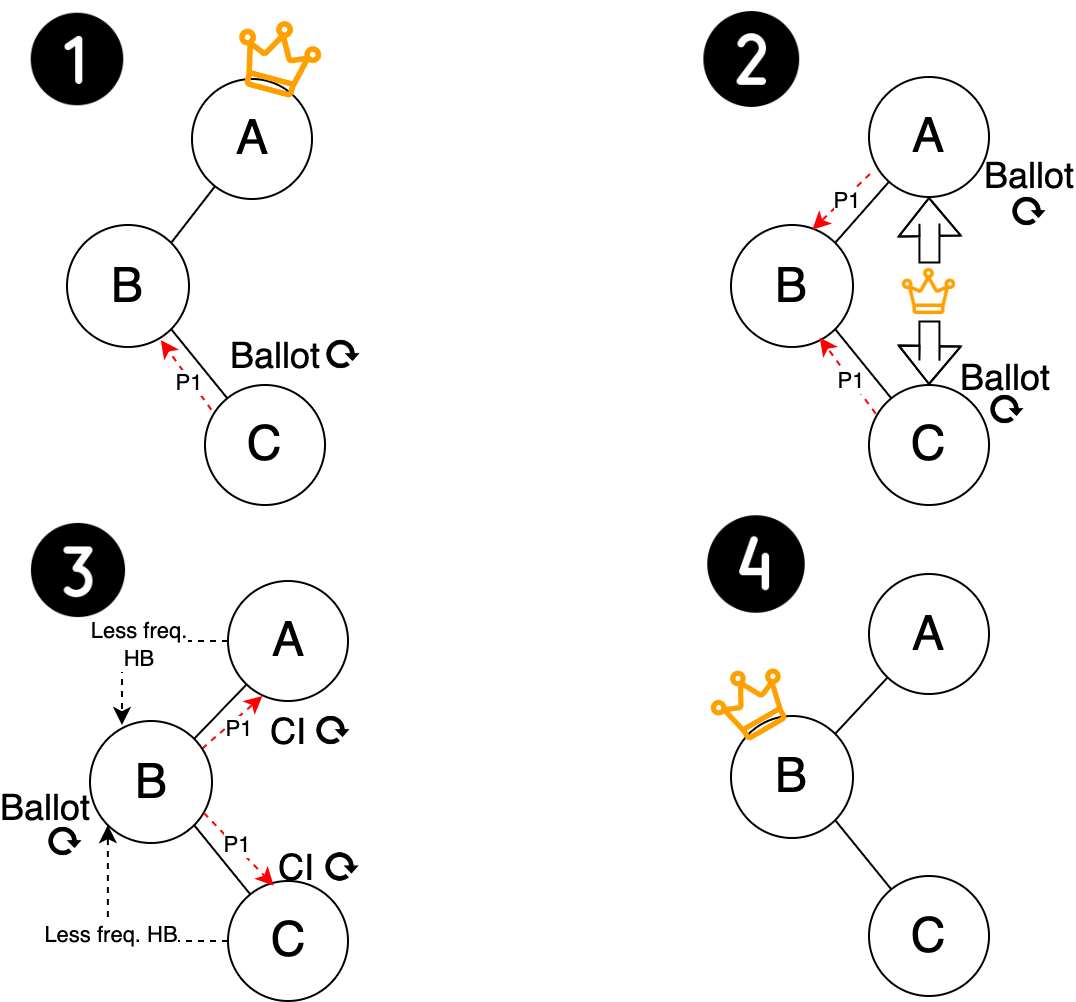}
    \caption{The leader churning partition, where Peer A and C disconnect from each other. 1. Peer C triggers timeout and prompts a leader election. Peer B becomes a follower of C due to a higher ballot number. 2. Peer A learns a new leader from B and becomes a follower, but it will start another election due to no heartbeats. A and C repeat leader elections, thus resulting low availability due to leadership churning. 3. B starts leader election as both A and C increase \textit{\textbf{commit\_interval}} and reduce the frequency of heartbeats. 4. describes that our MultiPaxos elects the stable Peer B as the leader to restore availability.}
    \label{fig:chained-partition-example}
\end{figure}

\autoref{fig:leader-loss-quorum-example} shows a scenario in a cluster of five fully connected servers, with Peer B as the original leader. 
In this scenario, a partition arises when all peers lose mutual connections, except for Peer E. As a result, Peer E is the only Peer that has quorum connections, while Peer B loses the quorum and becomes nonfunctional. 
Subsequently, Peer A, C, and D start leader elections but fail to secure sufficient votes to become the new leader. 
Previous studies suggest that the MultiPaxos fails under this partition as Peer E continues to respond to Peer B’s heartbeats and would not start the leader election~\cite{omnipaxos}. 
However, this claim is invalid and stems from an incorrect failure detector implementation in MultiPaxos~\cite{franken-paxos}.

In our MultiPaxos design, we demonstrate that a properly implemented heartbeat detector can facilitate the election of a new stable leader in such scenarios. 
When Peers A, C, or D trigger a leader election with a higher ballot and send Prepare requests to Peer E, Peer E updates its ballot number. 
This update causes Peer E to stop responding to heartbeats from the old leader, Peer B. In the meantime, no peers succeed in becoming the new leader, and Peer E does not receive heartbeats from any prospective new leader, leading to a timeout. 
This triggers another leader election initiated by Peer E. 
Ultimately, Peer E emerges as the new leader, restoring normal operation and allowing the system to resume progress.

\subsubsection{Leader Churning and Adaptive Timeout Setting}

While MultiPaxos generally exhibits resilience, it is not immune to all network partition scenarios. 
\autoref{fig:chained-partition-example} illustrates such a case in a three-server fully connected cluster, originally led by Peer A. 
When Peers A and C become unreachable from each other, leading to a partial partition, Peer C initiates a leader election and successfully becomes the new leader. 
However, when Peer A learns about C's new leadership through Peer B, it triggers another election because it does not receive heartbeats from C, resulting in a cyclic pattern of leadership change that causes a livelock situation within the cluster.

Though this problem draws a lot of attention \cite{omnipaxos, raft-examining}, we surprisingly discovered that a simple change to the failure detector could effectively resolve this issue, without any changes to the MultiPaxos algorithm. 
Originally, our failure detector utilized a random timer set to 2-2.5 times the \textit{\textbf{commit\_interval}}. 
We enhance this approach by introducing an adaptive timer mechanism. 
When a peer, such as Peer A or C, observes excessive leader elections initiated by itself within a certain timeframe, it increases its  \textit{\textbf{commit\_interval}}. 
This change leads to extended sleep periods before initiating another leader election, reducing the frequency of these elections. 

More importantly, the extended  \textit{\textbf{commit\_interval}} causes the temporary leader to send heartbeats less frequently, which increases the likelihood of a more stable peer, like Peer B, being elected as the new leader. 
Peer B's unchanged \textit{\textbf{commit\_interval}} results in a timeout quicker than Peers A and C, and it prompts to initiate a leader election. 
Since Peer B maintains connections with all other peers, it can effectively end the leadership churning once elected and become a stable leader. 
The cluster can return to normal availability quickly. Additionally, when the partition barrier dissolves, i.e. Peers A and C reestablish a connection, it does not disrupt Peer B’s current leadership, thereby maintaining the system’s overall availability.

\section{Evaluation}
In the evaluation of our MultiPaxos implementation, we aim to answer the following critical questions: 
(1) How does our MultiPaxos perform compared to other consensus protocol implementations? 
(2) How does our log-trimming approach perform in contrast to snapshotting and quorum-trim across various scenarios? 
(3) What level of resilience does our MultiPaxos exhibit in terms of throughput under various partial network partition scenarios?

Following our design specifications, we ported the pseudocode into four different languages. 
For simplicity, we opt for our Go implementation to compare with other open-source MultiPaxos or Raft projects, considering that a significant portion of such projects are implemented in Go. 
We employ gRPC~\cite{grpc} for inter-peer MultiPaxos communication and asynchronous TCP for client-to-server communication.

We conduct all experiments on both 3-node and 5-node clusters, utilizing AWS m5.2xlarge instances equipped with 8 vCPUs and 32\,GiB RAM. 
We analyze the results from the 3-node cluster, as similar trends across various experimental settings are observed in both configurations. 
The experimental setup includes a Zipfian request distribution, with key and value sizes set at 23 B and 500 B respectively~\cite{twitter1}. 
Prior to experiments, we populate the database with 1 million key-value pairs and initiate a 20-second warm-up phase. 
We run YCSB workload A (50\% reads and 50\% writes)~\cite{ycsb, ycsb-workloads}  on an independent m5.2xlarge instance employing 64 client threads. The reported results represent one sample obtained from five independent runs.

\subsection{Throughput vs. Latency}

We first examine the general performance of our MultiPaxos implementation in a common scenario. In this experiment, we compare the performance of MultiPaxos to several established SMR implementations, including Paxi from academic projects as well as production-ready systems like etcd and TiKV, under YCSB workload A for 3 minutes. 
As the focus of this paper is not on performance enhancement, the primary objective of this evaluation is to demonstrate that our MultiPaxos implementation offers reasonable and competitive performance when compared to these well-established systems. 
To ensure a fair comparison, we configure etcd and TiKV, which by default persist every operation entry in their logs, to use a RAM disk for this experiment. 

\begin{figure}
    \centering
    \includegraphics[width=\linewidth]{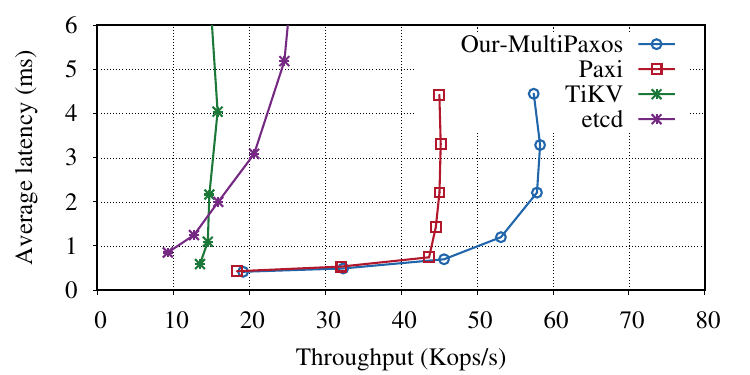}
    \caption{Throughput vs. Latency in 4 different SMR implementations. The data points correspond to 8, 16, 32, 64, 128, 192, and 256 concurrent clients. Data points with higher than 6\,ms latency are not shown.}
    \label{fig:throughput-latency}
\end{figure}

\begin{figure*}
    \centering
    \begin{subfigure}[t]{0.3\textwidth}
        \includegraphics[width=\textwidth]{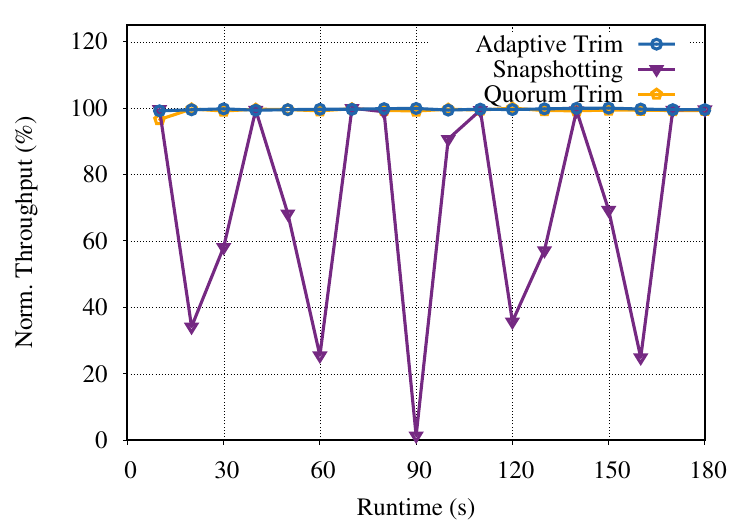}
        \caption{Norm. throughput - common path}
        \label{fig:common-path}
    \end{subfigure}
    \hfill
    \begin{subfigure}[t]{0.3\textwidth}
        \includegraphics[width=\textwidth]{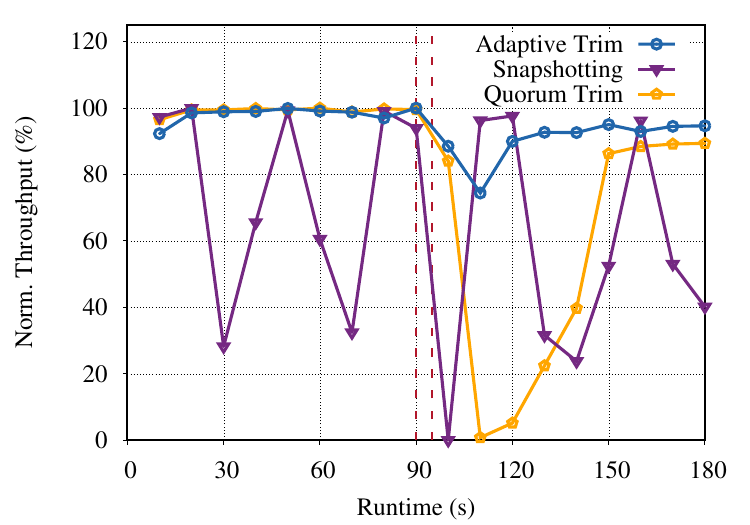}
        \caption{Norm. throughput - 5s disconnection}
        \label{fig:disconnect}
    \end{subfigure}
    \hfill
    \begin{subfigure}[t]{0.3\textwidth}
        \includegraphics[width=\textwidth]{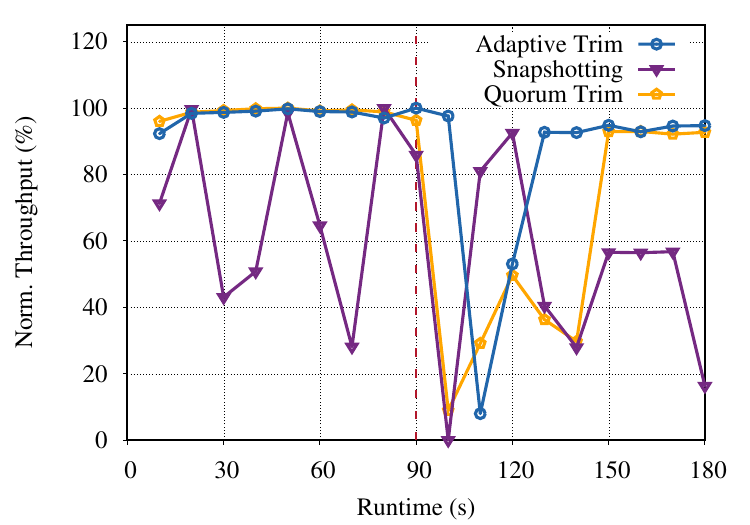}
        \caption{Norm. throughput - add a new node}
        \label{fig:join}
    \end{subfigure}
    \medskip
    \begin{subfigure}[5]{0.3\textwidth}
        \includegraphics[width=\textwidth]{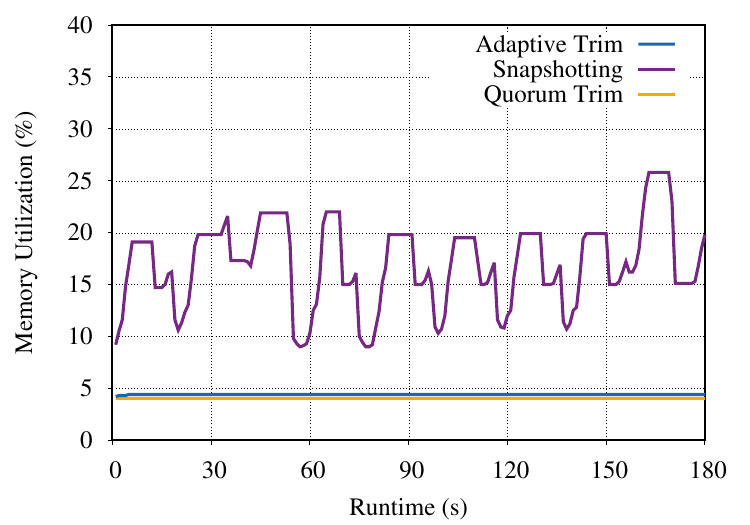}
        \caption{Mem. Utilization - common path}
        \label{fig:common-path-mem}
    \end{subfigure}
    \hfill
    \begin{subfigure}[5]{0.3\textwidth}
        \includegraphics[width=\textwidth]{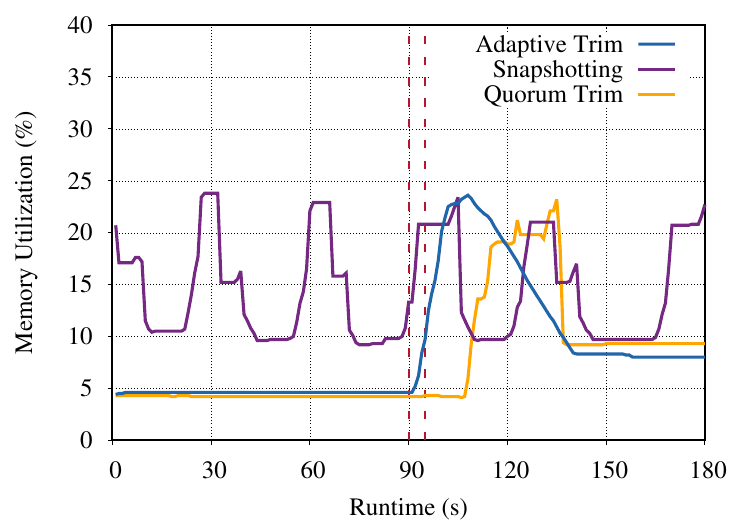}
        \caption{Mem. utilization - 5s disconnection}
        \label{fig:disconnect-mem}
    \end{subfigure}
    \hfill
    \begin{subfigure}[5]{0.3\textwidth}
        \includegraphics[width=\textwidth]{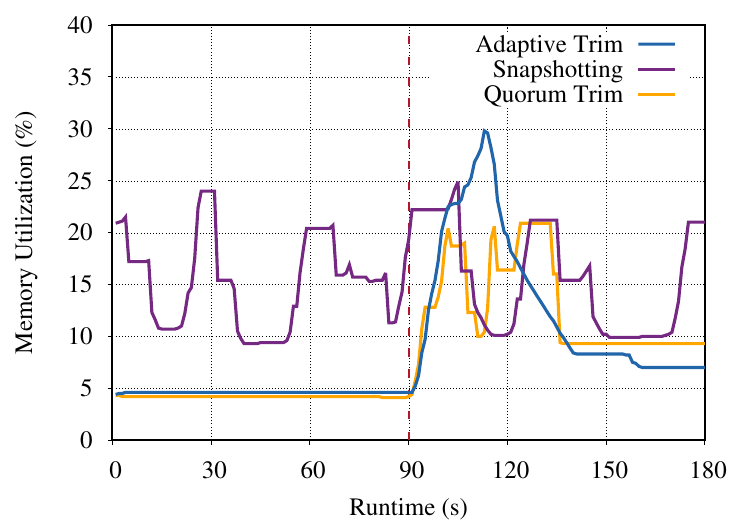}
        \caption{Mem. utilization - add a new node}
        \label{fig:join-mem}
    \end{subfigure}
    \caption{The impact of different log compaction methods. (a), (b), and (c) represent the normalized throughput regulated by their respective maximum throughput during the runtime. The two red dashed lines in (b) refer to the disconnection for 5 seconds, and the red dashed line in (c) stands for the event of a new node joining. Both events start at the 90th second.}
    \label{fig:log-compaction}
\end{figure*}

The results, as depicted in the \autoref{fig:throughput-latency}, reveal throughput and average latency metrics across a range of client numbers, from 8 to 256 closed-loop clients. 
While etcd and TiKV exhibit similar performance between 10-25\,Kop/s, MultiPaxos achieves nearly 60\,Kop/s. 
This difference in performance is expected, given the additional features incorporated in both production-ready systems, etcd and TiKV, which may impact their throughput.
In comparison to Paxi, MultiPaxos demonstrates similar throughput and average latency when the client count is below 32. 
However, as the number of clients increases, MultiPaxos reaches almost 1.5x maximum throughput while maintaining a similar average latency. 
This result indicates that our MultiPaxos excels in promptly processing operations with minimal delay, maintaining consistently low latency across a wide range of throughput levels. 
The latency of MultiPaxos starts to increase significantly only after the throughput exceeds 45\,Kop/s, suggesting a threshold at which the system begins to experience resource contention or processing bottlenecks.


\subsection{Log Compaction Overhead}

We next evaluate the impact of log compaction on performance and resource utilization, targeting three distinct scenarios: the common path, a temporary node disconnection, and the introduction of a new node. 
Notably, the effect observed when a node disconnects for a long period is similar to the case when a new node joins (with the need of sending snapshots instead of replaying log entries). Consequently, we only evaluate and analyze the scenario of adding a new node. 

We select the etcd’s Raft module as a snapshot-based approach, which is widely used by open source projects. For a fair comparison, we replace our MultiPaxos module with their consensus modules following their official example guidelines~\cite{etcd-raft-example}. 
The primary distinction lies in the decision of whether to take periodic snapshots. It ensures that the performance difference is not caused by other modules. 
For the snapshot interval, we set it to capture every 500K entries in the log. For the Quorum Trim approach, all instances are trimmed after execution, and the snapshotting happens on demand.

\autoref{fig:log-compaction} shows the real-time throughput and memory utilization throughout the runtime. 
As each implementation has different throughput limits, we normalize all throughput values by their respective maximum throughput, emphasizing the effect of snapshotting on throughput variability. 
Since the memory utilization remains consistent across all nodes, we exclusively present the memory utilization of the leader to save space. 
Whenever we add a new node or trigger a temporary disconnection, we perform these events at the 90th second, denoted by vertical red lines in the figures. Each peer disconnection lasts 5 seconds.

\autoref{fig:common-path} reveals that our Adaptive Trim and the Quorum Trim approach showcase a remarkably stable performance throughout the entire runtime in the common path, consistently achieving a maximum throughput. 
In contrast, the throughput of the snapshot-based approach exhibits a highly volatile pattern, with sharp peaks and downs, ranging from below 20\% to above 80\% of its maximum throughput. The variability underscores the considerable performance cost associated with snapshotting. 
Additionally, as shown in \autoref{fig:common-path-mem}, our approach also shows a low and stable level of memory utilization. 
The utilization is at the same level as the Quorum Trim, which removes all log entries immediately. 
This outcome is attributed to the consistent pace of trimming in-memory log entries in our method. 
Conversely, snapshotting not only leads to higher memory utilization but also exhibits notable fluctuations. 
These peaks in memory usage occur during snapshot generation, where substantial memory is temporarily allocated before being persisted to disk and eventually collected. 


However, we observe that snapshotting exhibits consistent performance in both the disconnection and new node scenarios, as shown in \autoref{fig:disconnect} and \autoref{fig:join}. 
This aligns with our expectations since the cost of snapshotting gets amortized throughout the runtime duration. 
Nevertheless,  transmitting snapshots consumes network bandwidth, occasionally resulting in multiple rounds of leader elections, potentially leading to continuing performance degradation even after the transmission of the snapshot, if a leader is not elected quickly, as shown in \autoref{fig:join} from 90th to 180th seconds.

\begin{figure*}
    \centering
    \begin{subfigure}[t]{0.45\linewidth}
        \includegraphics[width=\textwidth]{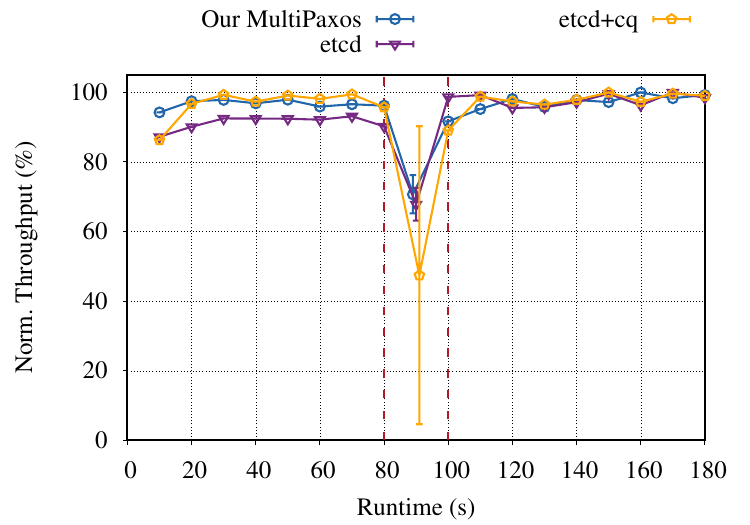}
        \caption{Normalized Throughput under leader-losing-quorum partition}
        \label{fig:5-node}
    \end{subfigure}
    \hfill
    \begin{subfigure}[t]{0.45\linewidth}
        \includegraphics[width=\textwidth]{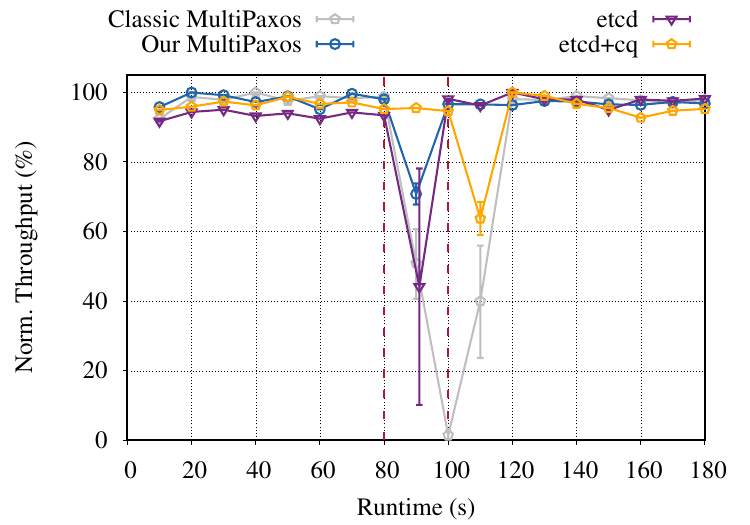}
        \caption{Normalized Throughput under chained-leader-churning partition}
        \label{fig:chained}
    \end{subfigure}
    \caption{The normalized throughput of different SMR implementations under partial network partitions. The error bars refer to the range of performance drops in multiple runs. The two dashed lines represent the start and the end of the partition. MultiPaxos refers to our MultiPaxos design with the adaptive timeout setting, while the Classic Multipaxos uses regular constant timeout; etcd is the integration of etcd's raft module with our key-value demo with CheckQuorum disabled; etcd+cq stands for etcd with CheckQuorum enabled.}
    \label{fig:enter-label}
\end{figure*}

In contrast, our log compaction mechanism is also influenced when such events occur, but any degradation is transient, and the case of disconnection has significantly less degradation than when a node joins. 
In the context of disconnection, the memory utilization spike arises from the suspension of log compaction, jumping from 5\% to nearly 25\%, and this increase persists for over 5 seconds (the disconnection duration), as shown in \autoref{fig:disconnect-mem}. 
This is because the re-connected follower requires several rounds of Commit messages to catch up on the progress and complete the batch commit. Subsequently, the leader can update the global\_last\_executed and start log trimming.

Regarding adding a new node, taking a snapshot of the current state machine is the only option in our Adaptive Trim method. 
Its cost is similar to snapshotting, encompassing the time needed for snapshot creation and transmission, along with Quorum memory allocation due to the substantial size of in-memory snapshots. 
Importantly, though Quorum Trim also relies on snapshotting, it almost doubles in duration to restore the performance than our Adaptive Trim approach, as depicted in \autoref{fig:join}.
When the peer is restored from the previous snapshot, it requires another new snapshot, as the leader keeps adding new entries and removes them immediately after execution. Thus, it results in more than one snapshot for restoring the status. 
Excessive snapshotting can lead to repeated leader elections as the leader spends a longer time on snapshotting.
Notably, all approaches do not restore to the maximum throughput after both events, which is related to the fail-slow follower~\cite{fail-slow}. 
In summary, even under scenarios of introduction or re-connection, the degradation of throughput with our log compaction mechanism is less than 10\% in most cases, while the periodic-snapshotting-based method may cause 40-60\% degradation during the runtime; our log compaction method also outperforms on memory utilization.

\subsection{Resilience under Partial Partition}


We next analyze the performance in scenarios involving partial network partitions. We specifically examine two types of partitions: the leader-losing-quorum and the chained-leader-churning scenarios. 
In both cases, we run the YCSB workload for 3 minutes, initiating the network partition at the 80th second and restoring the cluster at the 100th second, resulting in a 20-second duration. 
This duration is sufficient, as all clusters, regardless of the implementations, either regain availability within 20 seconds or remain unavailable indefinitely.

Similar to the previous experiment, our analysis primarily compares our MultiPaxos design to the etcd-integrated version. 
Additionally, to explore the impact of the failure detector on performance during network partitions, we include variants such as etcd with CheckQuorum enabled (namely etcd-cq) and MultiPaxos without adaptive timeout optimization (denoted as Classic MultiPaxos) in the comparison. 
The CheckQuorum option indicates that the leader steps down and allows a leader election if it loses connections to the majority of peers. 
Note that, in the etcd implementation, it also prevents the followers from answering any votes if they receive messages from the leader. 
Once again, we normalize throughput to the respective maximums of each implementation to ensure a fair comparison.

\autoref{fig:5-node} shows that the performance drop of MultiPaxos remains within the range of 65\% to 75\% of its maximum throughput during the leader-losing-quorum partition scenario. 
This decrease is primarily because the old leader loses quorum and halts progress during the partition. 
Upon partition initiation, the "stable peer," the only peer with connections to all others, only initiates a leader election after missing heartbeats, typically requiring at least two rounds of election timeouts (one for other followers initiating the election and another one for the stable peer starting its election). 
Once the stable leader is elected, throughput rebounds to the normal level, surpassing 95\%. 
Even after the partition is resolved, the throughput remains unaffected by changing network conditions, as the new leader maintains stable connections to all followers. 
Classic MultiPaxos exhibits similar resilience, so its data is omitted from the figure.

Etcd (without CheckQuorum enabled) performs comparably to MultiPaxos. During partition initiation, the stable peer updates its term number and disregards heartbeats from the old leader. 
This mirrors the behavior seen in MultiPaxos, where no peer becomes the leader until the stable one eventually initiates an election and assumes leadership. 
As a result, the unavailability period and performance decline in etcd is similar to MultiPaxos, falling within the 65\% to 70\% range.

However, when the CheckQuorum option is enabled, performance variance widens significantly. This variance results from the potential occurrence of multiple rounds of leader elections. 
The stable node ignores all vote requests if it has not triggered its election timeout. 
In some runs, the stable peer initiates leader election concurrently with other peers and secures leadership in a single round, achieving the upper bound of 90\%. 
On the other hand, the stable leader may delay its election, causing its term number to lag behind other candidates. Thus, it leads to multiple rounds of elections to catch up to the highest term number, resulting in a lower bound of less than 10\%.

\autoref{fig:chained} shows the normalized runtime throughput during the chained-leader-churning partition. 
As previously discussed, Classic MultiPaxos behaves as expected, with the cluster becoming almost entirely unavailable. 
Its throughput plummets by nearly 99\% and only begins to gradually recover after resolving the partition. 
In contrast, our MultiPaxos with optimized adaptive timeout settings demonstrates much greater resilience. The cluster experiences partial unavailability during leadership churning. 
The performance immediately resumes upon the election of a stable leader, even before the partition is resolved. Among the four implementations, our optimized MultiPaxos exhibits the best performance, with only a 30\% reduction in maximum throughput.

Interestingly, both etcd versions perform differently from the previous partition scenario. The range of performance degradation in etcd varies significantly, ranging from 20\% to 90\% reduction. 
This variation is attributed to the varied frequency of leader elections. When a follower loses connection with the leader, it repeatedly initiates leader elections with progressively higher term numbers. 
Although the stable peer does not grant votes, the repeated elections lead to a continuous rise in the term number, causing it to disregard heartbeats from the old leader. 
Consequently, the old leader also enters a cycle of repeated leader elections. Performance can only resume when the stable peer has an opportunity to initiate the election, resulting in varying degrees of performance drops. 

However, etcd+cq exhibits a different behavior. Its throughput remains unaffected during the partition, as the stable peer ignores all the votes from the disconnected follower and maintains the quorum. 
However, after the partition ends, the disconnected followers, now with higher term numbers, initiate leader elections, leading to several rounds of elections and snapshot transfers. This causes performance drops of approximately 30\% to 40\%. 
In summary, etcd+cq defers the cost until after the partition, while our MultiPaxos design elects a new stable leader during the partition, avoiding the case where a follower lags

\section{Related Work}
The original MultiPaxos paper, along with its variants introduced by Lamport, focused on algorithmic optimizations but omitted practical implementation details, posing challenges for real-world deployment~\cite{paxos-made-simple, fast-paxos, generalized-paxos, epaxos, flexible-paxos}. Subsequent research endeavors sought to address this gap, but their primary emphasis lay in specific implementation considerations and choices, rather than providing a comprehensive blueprint for reproduction~\cite{paxos-made-live, paxos-made-practical}. Although numerous academic research projects have produced open-source implementations, they often lack essential features crucial for practical applications, such as effective log compaction~\cite{nopaxos, paxi, skyros, libpaxos, franken-paxos, epaxos}. Additionally, these projects do not consider the coordination of different phases of the replication process.

In contrast, the industry has predominantly adopted the Raft consensus algorithm over MultiPaxos for building consensus modules~\cite{etcd, tikv, hashicorp, redis, rethinkdb, paxosstore, braft}. However, they are often language-limited, and detailed tutorials for integration are not always available. Furthermore, most industrial open-source frameworks heavily rely on snapshotting for log trimming. Notably, TiKV and etcd V3 use an alternative option, multi-version concurrency control, to avoid periodic snapshots.

While there have been studies on partial network partitions, they emphasize analysis instead of solutions~\cite{nifty, raft-examining, raft-consensus-reliability}. Other than that, OmniPaxos aims to enhance MultiPaxos' resilience under partial partitions but introduces complexity into the overall design and may not offer an optimal solution in certain types of partitions~\cite{omnipaxos}. Mutating ectd, designed for edge scenarios, comes at the cost of reduced consistency~\cite{mutating-etcd}. Nifty uses overlay networks to handle partial partitions~\cite{nifty}. 

\section{Conclusion}
In this paper, we bridge the critical gap between the neglected algorithmic details of MultiPaxos and its practical implementation. We provide the comprehensive and detailed design of MultiPaxos with step-by-step pseudocodes for all essential components, including leader election, log recovery, the commit phase, and the failure detector. These components, though critical to implementations, are overlooked in the algorithm descriptions. Our performance evaluation showcases that our MultiPaxos implementation exhibits competitive throughput and latency compared to established systems. We further introduce an innovative and lightweight approach to log compaction that eliminates the need for snapshots. Our results highlight its lower performance cost and reduced memory utilization, outperforming snapshotting in most cases. Additionally, we demonstrate the resilience of MultiPaxos under the partial network partitions and propose an adaptive timeout mechanism to enhance the MultiPaxos availability in the leader-churn scenarios. Our work not only addresses the gaps in MultiPaxos implementation but also offers practical tidbits to reduce runtime costs and reach better resilience.




\bibliographystyle{plain}
\bibliography{sample}

\appendix
\newpage
\null
\newpage
\onecolumn
\section{MultiPaxos Design Specifications}
\label{multipaxos-psudocode}

In this section, we will introduce the pseudocode of the MultiPaxos module. As the MultiPaxos module employs gRPC for inter-peer communication, we firstly present the specifications of gRPC messages used for the \textit{Commit}, \textit{Prepare}, and \textit{Accept} phases and the protobufs used by these messages. Though we provides the specification of the gRPC version, our design is not restricted by gRPC. Thus, we can replace protobufs with other types, e.g. bytes, JSON, etc., as long as they have the same fields in the message.

\begin{lstlisting}
- types/protobufs

    - commit_request protobuf
      - ballot_: ballot of the leader sending the commit RPC
      - last_executed_: last_executed_ of the leader sending the commit RPC
      - global_last_executed_: global_last_executed of the leader sending the
        commit RPC
      - sender: id of the sender

    - commit_response protobuf
      - type_: enum {ok, reject}
      - ballot_: ballot of the peer (valid only if type_ == reject)
      - last_executed_: last_executed_ of the follower responding to the commit
        RPC

    - prepare_request protobuf
      - ballot_: ballot of the sender
      - sender: id of the sender

    - prepare_response protobuf
      - type_: enum {ok, reject}
      - ballot_: ballot of the peer (valid only if type_ == reject)
      - instances_: Instances of the peer since global_last_executed_ (valid
        only if type_ == ok)

    - accept_request protobuf
      - instance_: Instance sent by the leader
      - sender_: id of the sender

    - accept_response protobuf
      - type_: enum {ok, reject}
      - ballot_: ballot of the peer (valid only if type_ == reject)

    - command protobuf
      - type_: enum {get, put, del}
      - key_: string
      - value_: string

    - instance protobuf
      - ballot_: ballot of the instance
      - index_: index of the instance in the log
      - client_id_: id of the client that issued the command
      - command_: command protobuf

    - result
      - type_: enum {ok, retry, someone_else_leader}
      - leader_: optional int
\end{lstlisting}

When running MultiPaxos phases, we send out requests to multiple peers and receive responses from them. For each phase we need to keep track of certain variables, such as the number of responses received and so on. Since we launch a separate thread for each request made to a peer, these threads need to share these variables and update them based on the response. We cannot make these variables class members because these thread in theory may outlive the MultiPaxos object (threadsanitizer complains). Therefore, for each phase we collect all the variables we need to keep track of in a single struct and allocate using a reference-counted pointer, and pass to each thread a reference-counted pointer to the struct. These structs are protected by a lock and each thread first acquires the lock before updating the struct. The last thread that finishes, deallocates the struct.

\begin{lstlisting}
- prepare_state
  - num_rpc_: number of RPC responses received
  - num_oks_: number of positive prepare responses
  - leader_: current leader
  - last_index_: highest index observed in instances received
  - log_: the merged log of positive responses
  - mu_: mutex protecting this struct
  - cv_: condition variable that each thread will signal upon completion
    (run_prepare_phase() sleeps on this and wakes up when a thread signals
    it and checks to see if quorum is reached, e.g.)

- accept_state
  - num_rpc_: number of RPC responses received
  - num_oks_: number of positive prepare responses
  - leader_: current leader
  - mu_: mutex protecting this struct
  - cv_: condition variable that each thread will signal upon completion
    (run_accept_phase() sleeps on this and wakes up when a thread signals it
    and checks to see if quorum is reached, e.g).

- commit_state
  - num_rpc_: number of RPC responses received
  - num_oks_: number of positive prepare responses
  - leader_: current leader
  - min_last_executed: the smallest last_executed of positive responses
  - mu_: mutex protecting this struct
  - cv_: condition variable that each thread will signal upon completion
    (run_commit_phase() sleeps on this and wakes up when a thread signals it
    and checks to see if all peers have responded, e.g).
\end{lstlisting}

The specification of the MultiPaxos object is listed as follows, including its members, constants, non-member functions, and methods. All variables and methods contains detailed introduction about how it works.

\begin{lstlisting}
- MultiPaxos

  - members
    - ballot_: int64_t; current ballot number known to the peer;
      initialized |max_num_peers_|, which indicates that there is no current
      leader, since valid leader ids are in [0, max_num_peers_). it is a 64-bit
      integer, the lower 8 bits of which is always equal to |id_| of the peer
      that chose |ballot_|, and the higher bits represent the round number. we
      preserve 8 bits for |id_| but limit |id_| to 4 bits to avoid an overflow
      when identifying a leader. initialized to |id_|. at any moment, we can
      look at the lower 8 bits of |ballot_| to determine current leader.

    - log_: non-owning pointer to Log instance.

    - id_: int64; identifier of this peer. initialized from the configuration
      file. currently, we limit the number of peers to 16; therefore, id_ is a
      value in [0, 16).

    - commit_received_: atomic<bool>; indicates whether a commit message was
      received during commit_interval_.

    - commit_interval_: milliseconds; time between sending consecutive commits.
      initialized from the configuration file.

    - port_: string; the port where the RPC server is listening

    - num_peers_: number of peers; initialized from the configuration file.

    - rpc_peers_: an array of RPC endpoints to peers, including to self.
      initialized from the configuration file.

    - mu_: mutex; MultiPaxos is a concurrent object -- multiple threads may
      concurrently call its |replicate| method. |mu_| protects shared data in
      MultiPaxos.

    - tp_: thread_pool; a thread pool to which we post RPC requests to peers;
      initialized from the configuration file.

    - cv_leader_: condition variable; commit_thread sleeps on this condition
      variable to get notified when this peer becomes a leader.

    - cv_follower_: condition variable; prepare_thread sleeps on this condition
      variable to get notified when this peer becomes a follower.

    - rpc_server_: a RPC server for handling incoming RPC requests; initialized
      based on the configuration file

    - rpc_server_running_: bool; a flag that indicates if the RPC server is
      running

    - rpc_server_running_cv_: condition variable; to avoid calling gRPC's
      shutdown() on the RPC server before it is started, we sleep on this
      condition variable in the stop_rpc_server() method; in the
      start_rpc_server() method, we signal this condition variable after
      starting the RPC server.

    - rpc_server_thread_: thread; runs the rpc server until shutdown() is
      called.

    - prepare_thread_running_: atomic<bool>; a flag that indicates if the
      prepare thread is running

    - prepare_thread_: thread; runs the function prepare_thread() until
      shutdown() is called.

    - commit_thread_running_: atomic<bool>; a flag that indicates if the commit
      thread is running

    - commit_thread_: thread; runs the function commit_thread() until shutdown()
      is called.

  - constants

    - id_bits_ = 0xff: the lower bits of |ballot_| we use for storing the id of
      the current leader.

    - round_increment_ = id_bits_ + 1: we add this constant to |ballot_| to
      increment the round portion of |ballot_| by one.

    - max_num_peers_ = 0xf: maximum number of peers.

  - non-member functions

    - extract_leader_id(ballot: int) -> int
      # returns the id embedded in the ballot
      return ballot & id_bits_

    - is_leader(ballot: int, id: int) -> bool
      # returns true if the peer id embedded in the ballot is the same as the id
      # parameter and false otherwise.
      return extract_leader_id(ballot_) == id_

    - is_someone_else_leader(ballot: int, id: int) -> bool
      # returns true if there is some leader and that leader's id is different
      # that the id parameter and false otherwise.

  - methods

    - constructor(log: *Log, cfg: config)
      ballot_ = max_num_peers_
      log_ = log
      id_ = config["id"]
      commit_received_ = false
      commit_interval = config["commit_interval"]
      port_ = config["peers"][id_]
      rpc_server_running_ = false
      prepare_thread_running_ = false
      commit_thread_running_ = false
      instantiate RPC stubs to each peer including self

    - start()
      start_prepare_thread()
      start_commit_thread()
      start_rpc_server()

    - stop()
      stop_rpc_server()
      stop_prepare_thread()
      stop_commit_thread()
      tp_.join()

    - start_rpc_server()
      # builds the RPC server and assigns it to rpc_server_, sets
      # rpc_server_running_ to true and signals rpc_server_running_cv_ in case
      # stop_rpc_server() was called first; starts the RPC server by calling
      # gRPC wait method in a separate thread so that this thread can continue
      rpc_server_ = build rpc server
      mu_.lock()
      rpc_server_running_ = true
      rpc_server_running_cv_.notify_one()
      mu_.unlock()
      rpc_server_thread_ = new thread ( rpc_server->wait() )

    - stop_rpc_server()
      # waits until the RPC server is started and then calls shutdown() on the
      # rpc server handle.
      mu_.lock()
      while !rpc_server_running_
        rpc_server_running_cv_.wait(mu_)
      rpc_server_.shutdown()
      rpc_server_thread_.join()

    - start_prepare_thread()
      # starts a thread, called prepare_thread_ to run the prepare_thread()
      # function
      prepare_thread_running_ = true
      prepare_thread_ = new thread ( prepare_thread() )

    - stop_prepare_thread()
      # stops prepare_thread_. first sets prepare_thread_running_ to false and
      # then notifies cv_follower_ because prepare_thread() function may be
      # asleep on it.
      mu_.lock()
      prepare_thread_running_ = false
      cv_follower_.notify_one()
      mu_.unlock()
      prepare_thread_.join()

    - start_commit_thread()
      # starts a thread, called commit_thread_ to run the commit_thread()
      # function
      commit_thread_running_ = true
      commit_thread_ = new thread ( commit_thread() )

    - stop_commit_thread()
      # stops commit_thread_. first sets commit_thread_running_ to false and
      # then notifies cv_leader_ because commit_thread() function may be asleep
      # on it.
      mu_.lock()
      commit_thread_running_ = false
      cv_leader_.notify_one()
      mu_.unlock()
      commit_thread_.join()

    - id() -> int
      # return the id of this peer; used only in unit tests
      return id_

    - ballot() -> int
      # return the current ballot
      acquire mu_ and release on exit
      return ballot_

    - next_ballot() -> int
      # description: gets the next ballot number by incrementing the round
      # portion of |ballot_| by |round_increment_| and setting the |id_| bits to
      # the id if this peer, since |ballot_| could have been generated by
      # another peer.
      mu_.lock()
      next_ballot = ballot_
      next_ballot += round_increment_
      next_ballot = (next_ballot & ~id_bits_) | id_
      mu_.unlock()
      return next_ballot

    - become_leader(new_ballot: int, new_last_index: int)
      # this function is called at the end of a successful prepare phase, once
      # we have received prepare responses from the quroum. it sets the ballot_
      # to the new ballot number (its first argument), and it also sets the last
      # index of its log to the highest index number observed in the instances
      # that it has received as a response to the prepare request (its second
      # argument). finally, it signals the cv_leader_ condition variable that
      # wakes up commit_thread_, which starts sending out commit messages to
      # supress leader election.
      mu_.lock()
      ballot_ = new_ballot
      log_->set_last_index(new_last_index)
      cv_leader_.notify_one()
      mu_.unlock()

    - become_follower(new_ballot: int)
      # this function is called when we receive a message from another peer with
      # a higher ballot number. this can happen in two places: (1) when we get a
      # response to an RPC that we sent out when running one of the three phases
      # (i.e. in one of run_prepare_phase, run_accept_phase, or
      # run_commit_phase) or (2) when we handle an RPC request from another peer
      # that runs one of the three phases (i.e. RPC handler for each message
      # type). in both scenarios, we already hold the global mu_ lock when we
      # call this function, which is why, unlike become_leader, it does not
      # require acquiring the lock. it sets this peers ballot to its argument,
      # and it checks to see if this peer just became a follower and if so, it
      # signals cv_follower_ condition variable to wake up prepare_thread_,
      # which starts waiting for commit messages from the new leader.
      old_leader_id = extract_leader_id(ballot_)
      new_leader_id = extract_leader_id(new_ballot)
      if new_leader_id != id_ || old_leader_id == max_num_peers_:
        cv_follower_.notify_one()
      ballot_ = new_ballot

    - sleep_for_commit_interval()
      # sleeps for commit_interval_, which was retrieved from the configuration
      # file; this is called by the commit thread to sleep between sending
      # commit messages.
      sleep(commit_interval_)

    - sleep_for_random_interval()
      # sleeps for a random interval that is chosen randomly to be in the range
      # of [1.5 * commit_interval_, 2 * commit_interval_]; this is called by the
      # prepare thread to sleep between checking if it has received a commit
      # message from the leader
      ci = commit_interval_
      sleep(ci + ci / 2 + rand(0, ci / 2))

    - received_commit() -> bool
      # returns true if commit_received_ has changed since this function was
      # last called. commit_received_ is set to true by the commit RPC handler
      # upon receiving a commit message; this function is called by the prepare
      # handler after sleeping for random amount (per
      # sleep_for_random_interval() function) to see if a new commit message was
      # received.
      return commit_received_.exchange(false)

    - replicate(cmd: command, client-id: int) -> result
      # this is the main entry point of the multipaxos object. it is called for
      # each client request received at the peer and can be called concurrently.
      # in the common case, this function runs the accept phase and returns ok
      # enum if the command is committed.
      ballot = ballot()
      if is_leader(ballot, id_):
        return run_accept_phase(ballot, log_->advance_last_index(), command, client-id)
      if is_someone_else_leader(ballot, id_):
        return result {type_: someone_else_leader, leader_: extract_leader_id(ballot)}
      # the first ever election is in progress; this only happens when the peers
      # first boot and there is no elected leader for some random period of
      # time, until a new leader emerges. once a leader is elected, there is
      # always some peer that is the leader.
      return result{type_: retry, leader_: N/A}

    - prepare_thread()
      # this function is run by the prepare_thread_ thread. it is a loop that
      # sleeps until this peer becomes a follower and then within this loop
      # starts another loop which keeps running the prepare phase until a leader
      # emerges. if this peer becomes a leader, the inner loop is exited and the
      # outer loop goes back to sleeping until becoming a follower. if some
      # other peer becomes a leader, then the inner loop keeps running and
      # checking that commit messages (heartbeats) are regularly received from
      # the current leader.
      while prepare_thread_running_:
        # wait until this peer becomes a follower; whenever we are woken up we
        # also check that the prepare_thread_running_ flag is still set in
        # addition to checking if we are still a leader so that the thread can
        # be gracefully shut down by setting the prepare_thread_running_ flag to
        # false.
        mu_.lock()
        while prepare_thread_running_ && is_leader(ballot_, id_):
          cv_follower_.wait(mu_)
        mu_.unlock()
        # at this point, we are a follower
        while prepare_thread_running_:
          sleep_for_random_interval()
          if received_commit():
            continue
          # we haven't received a commit so we should start leader election
          next_ballot = next_ballot()
          r = run_prepare_phase(next_ballot)
          if r:
            # prepare phase succeeded and we got back the highest index of
            # instances seen in prepare responses and a merged log of all
            # responses.
            [last_index, log] = *r
            become_leader(next_ballot, last_index)
            replay(next_ballot, log)
            break

    - commit_thread()
      # this function is run by the commit_thread_ thread. it has a similar
      # structure to the prepare_thread() function. it is a loop that sleeps
      # until this peer becomes a leader and then within this loop starts
      # another loop which keeps running the commit phase until it becomes a
      # follower again. when it becomes a follower, the inner loop is exited,
      # and the outer loop goes back to sleeping until becoming a leader.
      #
      # this function serves three purposes: (1) once it becomes a leader, it
      # keeps sending commit messages, acting as a heartbeat and letting other
      # peers know that it is still an active leader; (2) it also communicates
      # to other peers the instances that were committed at the leader, so that
      # the other peers could commit those instances as well; (3) it also helps
      # to advance global_last_executed_ and trim the log; to this end, it
      # receives last_executed_ from all peers, computes the minimum of these
      # and sends out the new global_last_executed_ at each iteration.
      while commit_thread_running_:
        # wait until this peer becomes a leader; whenever we are woken up we
        # also check that the commit_thread_running_ flag is still set in
        # addition to checking that we are still a follower so that the thread
        # can be gracefully shut down by setting the commit_thread_running_ flag
        # to false.
        mu_.lock()
        while commit_thread_running_ && !is_leader(ballot_, id):
          cv_leader_.wait(mu_)
        mu_.unlock()
        # at this point, we are a leader. before entering the loop sends commit
        # messages, get global_last_execute_.
        gle = log_->global_last_executed()
        while commit_thread_running_:
          ballot = ballot()
          # if we are not a leader any more, exit the inner loop and go back to
          # the outer loop and wait until we become a leader again.
          if !is_leader(ballot, id_):
            break;
          # run the commit phase, obtain the new global_last_executed for the
          # next iteration.
          gle = run_commit_phase(ballot, gle)
          # sleep before sending commit messages again.
          sleep_for_commit_interval()

    - run_prepare_phase(ballot: int) -> optional pair{int, map{int -> instance}
      # runs the prepare phase using the ballot argument for the outgoing
      # prepare requests, and if successful, it returns the highest index seen
      # in among the received instances and a log that is the result of merging
      # all of the logs received from other peers.
      prepare_state state
      state.leader_ = id_

      prepare_request request
      request.set_sender(id_)
      request.set_ballot(ballot)

      # we've already sent an RPC to ourselves and we have appended the instance
      # to our log.
      mu_.lock()
      if ballot > ballot_:
        state.num_rpcs++
        state.num_oks++
        state.log_ = log_.log()
        state.last_index = log_.last_index()
      else:
        return nullopt
      mu_.unlock()

      # for each peer run a closure in a separate thread that sends out an RPC
      # request and updates the state and signals the main thread.
      for peer in rpc_peers_:
        if peer.id == id_:
          continue
        tp_.post(lambda {
          # call RPC stub
          prepare_response response
          peer->prepare(request, response)
          state.mu_.lock()
          state.num_rpcs_++
          if response.ok():
            state.num_oks++
            for instance in response.instances:
              state.last_index = max(state.last_index, instance.index)
              log::insert(state.log_, instance)
          else:
            mu_.lock()
            if response.ballot > ballot_:
              become_follower(response.ballot)
              state.leader_ = extract_leader_id(ballot_)
            mu_.unlock()
          state.mu_.unlock()
          state.cv_.notify_one()

       # the main thread sleeps on the condition state.cv_ condition variable
       # and woken up each time a thread completes and checks to see if an
       # important condition has changed
       state.mu_.lock()
       # sleep while this peer is still a leader and quorum is not reached and
       # not all peers have responded; since we send each RPC to ourselves as
       # well, our condition check is <= and not <.
       while state.leader_ == id &&
         state.num_oks <= num_peers_ / 2 &&
           state.num_rpcs != num_peers_:
         state.cv_.wait(mu_)
       # we exited the loop meaning one of the conditions has changed
       # if we reached the quorum, respond with the highest index observed and
       # the merged log.
       if state.num_oks > num_peers_ / 2:
         return {state.last_index_, state.log_}
       # we haven't reached the quorum and have exited the while loop, so either
       # someone else became a leader or we didn't get a positive response from
       # the quorum, which means prepare phase failed, and thus we return a null
       # value for the optional
       return nullopt
       state.mu_.unlock()

    - run_accept_phase(ballot: int, index: int, cmd: command, client_id: int)
      accept_state state
      state.leader_ = id_

      instance instance
      instance.set_ballot(ballot)
      instance.set_index(index)
      instance.set_client_id(client_id)
      instance.command = command

      accept_request request
      request.set_sender(id_)
      request.instance = instance

      # we've already sent an RPC to ourselves and we have appended the instance
      # to our log.
      mu_.lock()
      current_leader = ExtractLeaderId(ballot_)
      if current_leader == id_:
        state.num_rpcs++
        state.num_oks++
        log_.append(instance)
      else:
        return {someone_else_leader, current_leader}
      mu_.unlock()

      for peer in rpc_peers {
        if peer.id == id_:
          continue
        tp_.post(lambda {
          accept_response response
          peer->accept(request, response)
          state.mu_.lock()
          state.num_rpcs_++
          if response.ok():
            state.num_oks++
          else:
            mu_.lock()
            if response.ballot > ballot_:
              become_follower(response.ballot)
              state->leader_ = ExtractLeaderId(ballot_)
            mu_.unlock()
          state.mu_.unlock()
          state.cv_.notify_one()

      state.mu_lock()
      while state.leader_ == id &&
          state.num_oks_ <= num_peers_ / 2 &&
          state.num_rpcs_ != num_peers_
        state.cv_.wait(mu_)
      if state.num_oks > num_peers_ / 2):
        log_.commit(index)
        return {ok, nullopt}
      if state->leader_ != id_:
        return {someone_else_leader, state.leader_}
      return {retry, nullopt}

    - run_commit_phase(ballot: int, global_last_executed: int)
      commit_state state
      state.leader_ = id_

      commit_request request
      request.set_ballot(ballot)
      request.set_sender(id_)
      request.set_last_executed(state.min_last_executed)
      request.set_global_last_executed(global_last_executed)

      state.num_rpcs_++
      state.num_oks++
      state.min_last_executed = log_.last_executed()
      log_.trim_until(global_last_executed)

      for peer in rpc_peers {
        if peer.id == id_:
          continue
        tp_.post(lambda {
          commit_response response
          peer->commit(request, response)
          state.mu_.lock()
          state.num_rpcs_++
          if response.ok():
            state.num_oks++
            if response.last_executed < state.min_last_executed:
              state.min_last_executed = response.last_executed
          else:
            mu_.lock()
            if response.ballot > ballot_:
              become_follower(response.ballot)
              state->leader_ = ExtractLeaderId(ballot_)
            mu_.unlock()
          state.mu_.unlock()
          state.cv_.notify_one()

      state.mu_lock()
      while state.leader_ == id && state.num_rpcs_ != num_peers_:
        state.cv_.wait(mu_)
      if state.num_oks == num_peers_:
        return state.min_last_executed
      return global_last_executed

    - replay(ballot: int, log: map{int -> instance})

    - prepare(request: prepare_request, response: prepare_response)
      # prepare RPC handler
      mu_.lock() # unlocks on return
      if request.ballot_ >= ballot_
        ballot_ = request_.ballot_
        return prepare_rpc_response{type_: ok,
                                    ballot_: N/A,
                                    log_: log_.instances()}
      # reject stale RPC requests
      return prepare_rpc_response{type_: reject, ballot_: ballot_, log_: N/A}

    - accept(request: accept_request, response: accept_response)
      # accept RPC handler
      if request.ballot_ >= ballot_:
        ballot_ = request.ballot_
        log_.append(request.instance_)
        return accept_rpc_response{type_: ok, ballot_: N/A}
      # stale message
      return accept_rpc_response{type_: reject, ballot_: ballot_}

    - commit(request: commit_request, response: commit_response)
      # description: handler for the commit RPC

      mu_.lock() # unlocks on return
      if request.ballot_ >= ballot_
        last_commit_ = time::now()
        ballot_ = request.ballot_
        stale_rpc = false
        log_.commit_until(request_.last_executed_, ballot_)
        log_.trim_until(request_.global_last_executed)
      return commit_rpc_response{last_executed_: log_.last_executed()}
\end{lstlisting}

\section{Log Design Specifications}
\label{log-psudocode}

Similarly, the specification of the Log object is listed as follows, including its members, constants, non-member functions, and methods. All variables and methods contains detailed introduction.

We can think of Log as an unbounded producer-consumer queue. From this perspective, the execute method below acts as the consume method of the queue, and the commit method below acts as the produce method of the queue. Technically, instances are inserted into the queue via the append method; however, they do not become executable until they are committed by calling commit on the instance. Wake-up happens only one way: the thread that commits wakes up the executor thread who executes instances by calling the execute method.

Since the executor thread may block on calling the execute method of log, we need to wake up the executor when we want to cleanly shut down the system. To this end, the log has a running flag and a stop method that sets the running flag to false and signals the condition variable upon which the executor thread is sleeping (due to calling the execute method). The execute method will return a nullopt after the log is stopped and the executor method can exit upon receiving nullopt.

Note that, we added assertions in the specification to validate whether invariants hold. Given that (1) the instances in the log must be executed in order, (2) \textbf{\textit{last\_executed}} is the index of the last executed instance, and (3) \textbf{\textit{global\_last\_executed}} is the index of the last instance that was executed in all peers, our log has the following invariants:

\begin{itemize}
    \item (i1) \textbf{\textit{global\_last\_executed}} <= \textbf{\textit{last\_executed}}.
    \item (i2) all instances in [\textbf{\textit{global\_last\_executed}}, \textbf{\textit{last\_executed}}] are executed.
    \item (i3) there is no executed instance after \textbf{\textit{last\_executed}}.
    \item (i4) there are no instances at indices < \textbf{\textit{global\_last\_executed}}.
\end{itemize}

\begin{lstlisting}
- Log

  - types/protobufs:

    - protobuf Command: a key-value command. Fields of a Command are
      - type_: enum { get, put, del }
      - key_: string
      - value_: string (valid only if type_ == put)

    - protobuf Instance: an entry in the log. Fields of an Instance are:
      - ballot_: ballot of the instance
      - index_: index of the command in the log
      - client-id_: id of the client in the leader that issued the command
      - state_: enum { in-progress, committed, executed }
      - command_: Command

      Each Instance starts in the in-progress state, and changes to the
      committed state once quorum has voted for it, and changes to executed
      state once the executor thread has its command executed.

  - members

    - running_: boolean indicating whether the log is still running.

    - kv_store_: a reference to the key-value store used by the execute command
      for executing key-value commands from the log.

    - log_: a map from int to Instance.

    - last_index_: index of the highest-numbered instance in the log. the log
      starts at index 1; therefore, last_index_ is initialized to 0, signifying
      an empty log.

    - last_executed_: index of the last executed instance. initialized to 0.

    - global_last_executed_: index of the last executed instance on all peers
      known to this peer. initialized to 0.

    - mu_: the mutex of the Log object that needs to be acquired before the
      object is modified.

    - cv_executable_: the condition variable on which the execute method sleeps
      and which the commit method signals.

    - cv_committable_: the condition variable on which the commit method sleeps
      on rare occasions and the append method signals.

  - non-members functions

    - insert(log: *Log, instance: Instance) -> bool
      # description: inserts |instance| to |log| if possible, and returns true
      # if an instance was inserted to an empty slot;
      # also checks for safety violations.

      # case (1): |log| doesn't have an instance at instance.index_, in which
      # case we simply insert the instance to |log|.
      i = instance.index_
      if log_[i] == empty
        log_[i] = instance
        return true

      # case (2): |log| has a committed/executed instance at |instance.index|,
      # in which case insert should be a no-op, and the command in |log| should
      # match the command in |instance|.
      if log_[i].state_ == (committed or executed)
       assert(log[i].command_ == instance.command_) << "case2"
        return false

      # case (3): |log| has an in-progress instance at |instance.index|, in
      # which case insert should examine the ballot numbers of both instances.
      #
      # if |instance.ballot| > |log[i].ballot| insert replaces the instance
      # in |log| with |instance|.

      if instance.ballot > log[i].ballot
        log[i] = instance
        return false

      # if |instance.ballot| == |log[i].ballot| it must be the case that
      # |instance.command| == |log[i].command| and insert is a no-op.

      if instance.ballot_ == log[i].ballot
        assert(log[i].command_ == instance.command_) << "case3"

      # if |instance.ballot| < |log[i].ballot| insert ignores the instance
      # because it is stale.

      return false

  - public methods

    - constructor(kv_store: KVStore)
      running_ = true
      kv_store_ = kv_store
      log_ = empty map
      last_index_ = 0
      last_executed_ = 0
      global_last_executed_ = 0

    - last_executed(void) -> int
      acquire mu_ and release on exit
      return last_executed_

    - log(void) -> map from int to instance
      acquire mu_ and release on exit
      return log_

    - global_last_executed(void) -> int
      acquire mu_ and release on exit
      return global_last_executed_

    - advance_last_index(void) -> int
      acquire mu_ and release on exit
      return ++last_index_

    - set_last_index(last_index: int)
      acquire mu_ and release on exit
      last_index_ = last_index

    - last_index(void)
      acquire mu_ and release on exit
      return last_index_

    - stop()
      acquire mu_ and release on exit
      set running_ to false
      signal cv_executable_

    - append(instance: Instance)
      # description: appends an instance to the log; called from the
      # accept_handler which runs on a separate thread.

      mu_.lock()
      i = instance.index_
      if i <= global_last_executed_
        mu_.unlock()
        return

      if insert(&log_, instance)
        last_index_ = max(last_index_, i)
        cv_committable_.notify_all()

    - commit(index_: int)
      # sets the state of the instance at index to committed and possibly wakes
      # up the executor thread if the log is executable.

      # commit is called exclusively by the leader, after it sends out accept
      # to all peers, including itself, and receiving ok responses from the
      # quorum. the accept handler in every peer inserts an instance into the
      # log and then responds with an ok if successful. most of the time, the
      # leader will immediately receive an ok from its own rpc handler before
      # receiving ok from the remote peers; therefore, by the time it calls
      # commit, it is likely to have an instance at the index_, and calling
      # commit is fine. in the rare cases when the leader receives ok
      # responses from remote peers before itself, we wait on the condition
      # variable that will eventually be signalled by append called from
      # accept_handler.
      mu.lock()
      while log_[index] == empty
        cv_commitable_.wait()

      # when running prepare, we will ask peers, including ourselves, to send us
      # their log starting at their global_last_executed_ and merge those logs.
      # then we run accept on all instances in the merged log starting after
      # global_last_executed_. hence, we may run accept on an instance that is
      # already committed or even executed in our log. our accept handler will
      # not touch log_ for such instances but it will respond with an accept and
      # eventually, we may run commit for such instances, in which case we will
      # end up here. for those instances, commit must be a no-op. hence, we will
      # update an instance's state to committed only if it is in in-progress
      # state.
      if log_[index_].state == in-progress
        log_[index_].state = committed

      # we must do this check every time because it may be an instance that we
      # merged into our log from a remote peer that was already in committed
      # state. in this case, we should wake up the thread to execute the
      # instance on our state machine.
      if (executable())
        cv_executable_.notify_one()
      mu_.unlock()

    - execute() -> optional(client-id, result)
      # As described above, this method acts as a consume method of a
      # producer/consumer queue. Therefore, it sleeps until it is woken up by
      # someone calling commit (i.e. produce method of the queue). Once woken
      # up, it executes one instance, sets the state of the instance to
      # executed, increments last_executed_, and returns the result and the id
      # of the client that originated the command.
      mu_.lock()
      while running_ and not is_executable()
        cv_executable_.wait()

      if !running_:
        return nullopt

      instance = &log_[last_executed_+1]
      result = kv_store_.execute(instance.cmd)
      ++last_executed_
      return (instance.client-id_, result)

    - commit_until(leader_last_executed: int, ballot: int)
      # called from the heartbeat handler after receiving last_executed from the
      # leader (i.e. leader_last_executed). sets the state of all the instances
      # from last_executed_ until leader_last_executed and wakes up the executor
      # thread if necessary.
      #
      # since we send heartbeats to ourselves as well, commit_until will be
      # called on the leader as well, but it will most likely to have no effect
      # because the entries will be committed immediately after the leader
      # receives acks from the quorum.

      assert leader_last_executed >= 0
      assert ballot >= 0

      mu_.lock()
      for (int i = last_executed_ + 1; i <= leader_last_executed; ++i)
        # we may receive a heartbeat before we receive the accept message;
        # therefore, the heartbeat handler may run this function while there
        # is a gap in the log. when we see a gap, we break out of the loop and
        # try committing the next time we receive heartbeat from the leader;
        # hopefully, by that time, we will have received the accept message
        # and the gap will disappear.
        if (log_[i] == empty)
          break

        # |ballot| will determine whether we will commit instances in the log.
        # we can have three cases with respect to |ballot| and the ballot of
        # instances in the log after last_executed_:
        #
        # case (1) |ballot| is smaller: this is the impossible case because
        # commit_until is always called from the heartbeat_handler, which will
        # reject heartbeats with a lower ballot number than ours. we assert this
        # impossiblity below.
        assert(ballot >= log_[i].ballot_)

        # case (2) they are equal: this is the common case. as a follower, we
        # will usually have in-progress instances in our log; we will
        # receive |leader_last_executed| that is larger than last_executed_, and
        # we will catch up by committing instances in our own log.
        if (log_[i].ballot_ == ballot)
          log_[i].state = committed

        # case (3) |ballot| is larger: this is the uncommon case, and it can
        # happen as follows: (a) we experience a partition, (b) a new leader
        # emerges and establishes new commands for the instances that are
        # already in our log, and (b) we reconnect. in this scenario, we will
        # have stale commands in our log and we cannot commit them because it
        # will violate safety. in this scenario, we will do nothing and return,
        # and as a result, prevent global_last_executed_ from advancing, until a
        # new leader is elected, and that new leader replays every instance
        # since global_last_executed_ and we discover the new commands that we
        # missed during the partition.

      if (executable())
        cv_executable_.notify_one()
      mu_.unlock()

    - trim_until(leader_global_last_executed: int)
      # called from the heartbeat handler after receiving global_last_executed
      # from the leader (i.e. |leader_global_last_executed|). removes all
      # instances in [global_last_executed_+1, leader_global_last_executed] from
      # the log_.

      while global_last_executed_ < leader_global_last_executed
        ++global_last_executed_
        # case (1): the following assertion follows from the invariants (i1) and (i2)
        assert(log_[global_last_executed_].state == executed)
        del log_[global_last_executed_]

    - instances() -> Instance[]
      # return all instances in the log for a response to a prepare message;
      # since global_last_executed_ is the tail of the log and last_index_ is
      # the head of the log, we return all instances in-between.
      mu_.lock()
      instances = Instance[]
      for i = global_last_executed_ + 1; i <= last_index_; ++i
        if log_[i] != empty
          instances.append(log_[i])
      return instances

    - is_executable(void) -> bool
      # preconditions: mu_ must be held

      # returns true if the log contains an executable instance, i.e. the
      # instance right after last_executed_ is committed.
      return log_[last_executed_+1] != empty &&
        log_[last_executed_+1].state == committed

    # this method is used only in unit tests
    - at(index: int) -> pointer to instance
      return a constant pointer to the instance
\end{lstlisting}

\end{document}